\documentclass[10pt, conference, letterpaper]{IEEEtran}
\usepackage{graphicx}
\usepackage{amssymb, amsmath}
\usepackage{subfigure}
\usepackage[hyphens]{url}
\usepackage{epstopdf}
\usepackage{algorithm}
\usepackage{algpseudocode}
\usepackage{expl3}

\newcommand{\mycomment}[1]{}

\newcommand{\etal}{et al.\@ }

\begin{document}

\title{Adaptation Logic for HTTP Dynamic Adaptive Streaming using Geo-Predictive Crowdsourcing}


\author{
\IEEEauthorblockN{Ran, Dubin, Ofer Hadar,Itay Katz, Ori Mashiach} 
\IEEEauthorblockA{Communication Systems Engineering\\Ben-Gurion University of the Negev\\ Israel\\}  
\and
\IEEEauthorblockN{Amit Dvir} 
\IEEEauthorblockA{Center for Cyber Technologies\\Department of Computer Science\\ Ariel University\\ Israel\\} 
\and
\IEEEauthorblockN{Ofir Pele} 
\IEEEauthorblockA{Center for Cyber Technologies\\Department of Computer Science\\Department of Electrical and Electronics Engineering\\Ariel University\\ Israel\\} 
}

\maketitle

\begin{abstract}
The increasing demand for video streaming services with high Quality
of Experience (QoE) has prompted a lot of research on client-side
adaptation logic approaches. However, most algorithms use the
client's previous download experience and do not use a crowd knowledge
database generated by users of a professional service. We
propose a new crowd algorithm that maximizes the QoE. Additionally, we
show how crowd information can be integrated into existing algorithms
and illustrate this with two state-of-the-art algorithms. We evaluate
our algorithm and state-of-the-art algorithms (including our modified
algorithms) on a large, real-life crowdsourcing dataset that contains
336,551 samples on network performance. The dataset was provided by
WeFi LTD. Our new algorithm outperforms all other methods in terms of QoS (eMOS).
\end{abstract}

\section{Introduction}
\label{Introduction}

Dynamic Adaptive Streaming over HTTP (DASH) \cite{DASH_RFC_1} is
the HTTP Adaptive Streaming (HAS) standard. It has been recently
adopted by YouTube (Google) and Netflix. DASH splits a video into
chunks and encodes each into several quality representations.

A client's DASH application often has a smart Adaptation Logic (AL)
module. The AL module is responsible for selecting the most suitable
quality representation to enhance the client's Quality of Experience
(QoE) while considering factors such as the client's buffer and playback
delay. QoE is affected by factors such as the number of quality
changes and their sizes.

There is a tradeoff between increasing the video quality and
buffering additional video segments. A client's player often buffers a
high number of segments to overcome network outages. 

Most of the current AL methods
\cite{KLUDCP,mok2012qdash,DubinHD13,WuLR2014,proxyMuller,de2013elastic,muller2012evaluation,PIO,MAL_conf},
estimate the next suitable segment based on estimates of previous
segments without taking into account the future network
characteristics. However, knowledge of geo-location network conditions can
enable better decisions.

The term crowdsourcing was introduced by Howe
\cite{howe2009crowdsourcing}. Howe defined crowdsourcing as the
act of taking a task traditionally performed by a designated agent
(such as an employee or a contractor) and outsourcing it by making an
open call to an undefined but large group of people, especially from
an online community.

In the case of video adaptive streaming, crowdsourcing makes it
possible to collect mobile network data anonymously and
automatically. This is done using an application specially designed to
improve the AL decision. Neidhardt et
al. \cite{neidhardt2013estimating} reports that using many of the
existing open datasets leads to low accuracy because of extreme
outliers and few measurements for some of the cells. They note that
cellular location providers do not provide their complete data. In
this work, we present a real-world crowdsourcing dataset and test our
proposed solution based on different users.

We propose a Geo-Predictive Adaptive Logic (GPAL) algorithm based on
crowdsourcing data on network performance provided by WeFi.  WeFi
collects granular information on mobile network performance and
application usage from millions of devices, down to a $10 \times 10$
meter geographical resolution.  A short 20-record sample of these data
can be found in \cite{URL_OF_DATA}.

We dub our new crowd algorithm GPAL and show that it outperforms
state-of-the-art algorithms. Moreover, we show that existing
adaptation algorithms can be improved by using crowd services. However,
our algorithm outperforms these algorithms even in this scenario.  It
is worth noting that our crowd sourcing data was generated by users of a
professional service and not by a simulation.

The remainder of this paper is organized as follows: Section
\ref{Related Work} describes related work. Section \ref{Proposed
  Geo-Predictive Algorithms} presents our proposed crowd
algorithms. Section \ref{datasets} present our dataset
characteristics. Section \ref{Experiments and Results} presents the 
experimental setup and results. Section \ref{Conclusions} discusses 
future work and conclusions.

\section{Related Work}
\label{Related Work}
DASH AL is a well-investigated research topic.  AL research can be
roughly divided into two different groups: past estimation based AL
and crowdsourcing based AL. Most work has investigated
past estimation algorithms.

M{\"u}ller et al. \cite{muller2012evaluation} suggested a buffer based
decision algorithm that uses the previous segment bandwidth estimates
and the user's current buffer duration to select a suitable quality
representation for downloading. The Multicast Adaptation Logic (MAL)
algorithm \cite{MAL_conf} uses a double Exponential Moving Average
(EMA) algorithm. One smooths the buffer size estimate and the other
smooths the bandwidth estimate.  This is done to select the most
suitable segment. Although MAL was designed for multicast, it
achieves good performance in unicast networks \cite{MAL_conf}. 

Crowdsourcing AL methods have attracted much less attention than past
estimation based methods. Hung et al. \cite{huang2014lbvs} proposed a
video streaming control mechanism based on location to overcome signal
variations in train tunnels and underground areas. Geo-location
frameworks that have the ability to predict future network conditions
based on a \textit{bandwidth lookup service} and similar concepts can
be found in
\cite{curcio2010geo,singh2012predictive,yao2012improving,riiser2012video,hao2014gtube}.
Acharya et al.\cite{acharya2008congestion} evaluated rate-adaptation
in a vehicular network based on signal strength and throughput at a
location as an indicator for congestion.  Curcio et
al. \cite{curcio2010geo} and Singh et al. \cite{singh2012predictive}
suggested server-side prediction algorithms for RTP streaming. Curcio
et al. \cite{curcio2010geo} suggested a framework with a predictive
server which obtains: route, speed, location and throughput from the
client. However, this study was based on simulation rather than
real-world data.  Singh et al. \cite{singh2012predictive} proposed
building a Network Coverage Map Service (NCMS) to make rate-control
decisions over a Real Time Protocol (RTP) using server-side adaptation
algorithm. Singh et al. however did not investigate performance on
datasets with a higher geographical coverage or more diverse network
connectivity conditions.

Yao et al. \cite{yao2012improving} showed that past bandwidth
information is a good indicator of the actual bandwidth at a given
location. Yao et al. found that location had greater influence than
time, based on traces. Nevertheless, their performance evaluation did
not take into account the number of switches or the playout buffer
size.  Furthermore, it was gathered from a small set of vehicles.

Riiser et al. \cite{riiser2012video} proposed a buffer based and a
crowd-based algorithm. Riiser et al. concluded that using past
bandwidth lookups led to far fewer rebuffering events and stabler
quality. Han et al. \cite{han2013maserati} investigated the extent to
which the available user mobile channel bandwidth is affected by constraints including location, time, speed, humidity and
cellular network type (3G/4G). Their scheme, called MASERATI,
outperformed Pure-DASH and LoDASH, where Pure-DASH only uses the download
throughput and LoDASH uses location based bandwidth predictions
as in \cite{Eve11,riiser2012video}.

Liu et al. \cite{Liu_10} suggested comparing the segment fetch time
with the media duration contained in the segment to detect congestion
and probe the spare network capacity. Liu's algorithm uses
conservative step-wise up switching and aggressive down switching. Hao
et al.\cite{hao2014gtube} suggested two algorithms: 1-predict and
n-predict. The 1-predict algorithm uses the playout buffer and the
next prediction to determine the most suitable representation to
download. The n-predict algorithm uses the average throughput of the
next n time steps as the algorithm's current prediction. Hao et
al. \cite{hao2014gtube} evaluated Liu et al.'s algorithm and found
that it achieved stable video quality but with a very
low average bitrate. They showed that n-predict outperformed
Liu et al's algorithm as well as 1-predict .

Zou et al. \cite{Zou2015} demonstrated that leveraging bandwidth predictions
can significantly improve QoE. They designed an algorithm that
combines bitrate prediction and rate stabilization. They
showed that during startup, their algorithm had more than four times
better video quality than heuristic-based algorithms.

Riiser et al. \cite{riiser2013commute} recorded $3G$ mobile traces in
Oslo, Norway, while traveling in different types of public transportation
(metro, tram, train, bus and ferry). However, the number of
contributors in the dataset was small.

Table \ref{tab:comp_2} summarizes the papers presented above.

\begin{table*}
\caption{Comparison of Algorithms}
\begin{tabular}{|p{1.5cm}|p{1.2cm}|p{2cm}|p{1.5cm}|p{1.5cm}|p{1.5cm}|p{2cm}|p{1.5cm}|p{1.5cm}|}
\hline 
Paper  & 
Streaming Protocol & 
Idea  & 
Trigger & 
Action&Quality Adjustment & 
Compared Algorithms& 
Observed Metrics & 
Mobility Simulate 
\tabularnewline
\hline 
Singh et el. \cite{singh2012predictive} - Geo-location Assisted Streaming System
(GLASS) Rate-Switching &
RTP/UDP + Temporal Maximum
Media Stream Bit rate Request (TMMBR) & 
Avoiding buffer underrun - client looks ahead at locations in its
vicinity for bad coverage &
Future Coverage Hole &
Client Pre-Buffer &
Client media rate switch according to available throughput in the coverage hole & 
No adaptation (RTCP), rate switching GLASS, late scheduling GLASS, Omniscient (Optimal) &
packet loss rate, average receive rate, Y component of the PSNR, throughput & 
Actual specific bandwidth trace 
\tabularnewline
\hline 
Yao et al. \cite{yao2012improving} - BW-MAP-TFRC & 
adaptive TCP streaming with TCP Friendly
Rate Control (TFRC) \cite{handley2002tcp} & 
Avoiding packet loss - client updates the server when it changes its
location.  The server determines the average bandwidth at that
location in the past &
Location changed by client followed by a new BW value & 
Server changes its sending rate & 
Short freezing of the TFRC and disabling the normal operation of TFRC when needed & 
TFRC and BW-MAP-TFRC &
estimated Mean Opinion Score \cite{sg122005model}, Peak Signal-to-Noise Ratio (PSNR), Glitch (Drop
in the streaming quality) &
Actual specific bandwidth trace 
\tabularnewline
\hline 
Riiser et al. \cite{riiser2012video} & 
Apple Live HTTP & 
Minimizing rapid fluctuations in quality and avoiding buffer underrun
- client's estimate of the number of bytes that it can download during
the remaining time of the trip &
Client receives a sequence of bandwidth averages for its whole path & 
Client plans which quality levels to use & 
Apple Live HTTP mechanism & 
Buffer-Based Reactive, History-Based Prediction, Omniscient Prediction (Optimal) & 
Buffer size, selected representations & 
Actual specific bandwidth trace
\tabularnewline
\hline 
Han et al. \cite{han2013maserati} - MASERATI & 
DASH & 
Avoiding frequent or large video quality changes &
The algorithm finds the most similar database entry and estimates the
available bandwidth &
The bit rate of the next video segment is defined by that bandwidth &
DASH Adaptation mechanism &
Pure-DASH, Location-based DASH (LoDASH) \cite{riiser2012video}, MASERATI & 
Playout Success Rate, Quality of Segments, Frequency of Quality
Changes, Degree of Changed Quality Level &
Actual specific bandwidth trace 
\tabularnewline
\hline 
Hao et al.\cite{hao2014gtube} - 1-predict, n-predict & 
DASH & 
Achieving continuous playback - DASH Based algorithm with an
additional function to anticipate future path and bandwidth, and to
determine the predicted rate &
The server calculates the possible bandwidth and sends it to the client & 
DASH Client applies the best quality level it can afford & 
DASH Adaptation mechanism & 
Liu et al. \cite{Liu_10}, Adobes Open Source Media Framework (OSMF),
1-Predict, N-Predict &
Segment representation Level, Ratio of bandwidth utilization, rate
of video quality level shift &
Actual specific bandwidth trace 
\tabularnewline
\hline 
Zou et al. \cite{Zou2015} - PBA & 
DASH & 
Avoiding stalls, preserving stability while maintaining improved average
quality - the client decides which quality to pick using the buffer
state and the quality of historical chunks &
Buffer occupancy changes all the time during download & 
Client can decide when to download and quality level & 
DASH & 
FESTIVE \cite{jiang2012improving} , BBA \cite{huang2014buffer}, optimal( mixed integer linear programing), PBA & 
Average quality rate supplied in the first 360s/32s, Number of stalls,
Number of switches &
Actual specific bandwidth trace
\tabularnewline
\hline 
\end{tabular}
\label{tab:comp_2}
\end{table*}



\section{Proposed Geo-Predictive Algorithms}
\label{Proposed Geo-Predictive Algorithms}

We define the user playout buffer as $B(t)$. The goal of the AL modules is
to maximize the overall quality of the stream, while eliminating
rebuffering ($B(t) > 0$). We measure the quality in terms of its eMOS
score \cite{claeys2014design} as shown in
Eq. \ref{eq:Qualitymaximization2}.

\begin{align} 
\label{eq:Qualitymaximization2}
\begin{split}
 \max (\text{eMOS}) \;\;\; & s.t: \\
 \forall t > t_{start} \;\;  0 < B(t) & \leq B_{max}
\end{split}
\end{align} 

We first show how to integrate crowd information to existing
algorithms. This is done by estimating the bandwidth. The estimate is
based on the crowd and not on the network.  We demonstrate the
approach with two state-of-the-art algorithms, MAL (Section
\ref{Geo-MAL}) and MaxBW (Section \ref{Geo-MaxBW Adaptation Logic}).

We also present a novel Geo-Predictive Adaptation Logic (GPAL)
algorithm that is designed to maximize the QoE (Section
\ref{Geo-Predictive Adaptation Logic (GPAL)}).

Our crowd bandwidth estimation algorithm is presented in
Algorithm \ref{alg:algo0}.  This algorithm was used for GPAL,
Geo-MaxBW and Geo-MAL.

\begin{algorithm}
\caption{Geo predictive bandwidth estimation algorithm, used by GPAL, Geo-MaxBW and Geo-MAL.} 
\label{alg:algo0}
\begin{algorithmic}[1]
\State \textsl{g}: current mobile geo-location.
\State \textsl{v}: current mobile speed.
\State \textsl{w}: highest quality average file size.
\State \textsl{f}: last downloaded segment throughput estimate.
\State \textsl{radius}: search radius.
\State $X_{bw}(t)$: bandwidth estimate for the current time ($t$).
\State \textsl{estimate}: \textsl{g,v}
\State \textsl{seg} \;=\; $\frac{\textsl{v}\textsl{w}}{\textsl{f}}$
\State $X_{bw}(t)$ \; =\; \textsl{getCrowdPrediction(radius, g, seg)}
\end{algorithmic}
\end{algorithm}

\subsection{Geo-MAL}
\label{Geo-MAL}
Dubin \etal \cite{MAL_conf} showed that using a Double Exponential
Moving Average (DEMA) estimator achieved good results in unicast and
multicast networks. Based on these results, we present a new
Geo-predictive MAL algorithm using a crowdsource adapted DEMA
estimator (Eqs. \ref{eq:EMA_buffer}-\ref{eq:EMA_BW}). The full
algorithm is presented in Algorithm \ref{alg:algo1}.

The DEMA estimator uses a parameter to balance the influence of the
current measurement vs. the influence of the previous estimate on
the current estimate. Increasing the parameter increases the weight
of the current measurement and decreases the weight of the previous
estimate. In MAL \cite{MAL_conf}, they used $\alpha$ to denote the
parameter used for the client's buffer estimate and $\beta$ to denote
the parameter used for the channel bandwidth estimate. One of the main
challenges is to choose appropriate values for $\alpha$ and $\beta$ to
best comply with the requirements of
Eq. \ref{eq:Qualitymaximization2}. Similar to \cite{MAL_conf} we used
$\alpha = 0.2$ and $\beta = 0.08$. We define $S_{b}(t)$ as the
smoothed buffer estimate and $S_{bw}(t)$ as the smoothed bandwidth estimation:

\begin{align}
\label{eq:EMA_buffer}
\begin{split}
S_{b}(0) & = B(0) \\ 
\forall t > 0 \;\; S_{b}(t) & = (\alpha)B(t)  + (1-\alpha) S_{b}(t-1)\\
\end{split}
\end{align}

\begin{align}
\label{eq:EMA_BW}
\begin{split}
S_{bw}(0)   &= X_{bw}(0) \\ 
\forall t > 0 \;\; S_{bw}(t) &= (\beta) X_{bw}(t)  + (1-\beta) S_{bw}(t-1)\\
\end{split}
\end{align}

\begin{algorithm}
\caption{Geo-MAL: Geo Predictive MAL Algorithm} 
\label{alg:algo1}
\begin{algorithmic}[1]
\State \textsl{critical}: playout buffer contains 2 segments.
\State \textsl{low}: playout buffer contains 4 segments.
\State \textsl{$B_{max}$}: maximum buffer size.
\State \textsl{almost full}: playout buffer contains $B_{max} -2$ segments.
\State \textsl{safety factor}: 0.5.
\State \textsl{Estimate $X_{bw}(t)$ for each segment download.}
\If {start or re-buffering}	
	\If{next representation$<$estimated bandwidth $\cdot$ safety factor}
			\State Ask for the highest representation available under that condition
	\EndIf
\EndIf
\If{segment is received}
\If{
                \Big(buffer is depleting\Big) and \newline 
\phantom{aaaaa} \Big(\big(buffer level $\leq$ critical\big) or \newline
\phantom{aaaaaaa}    \big((buffer level $\leq$ low) and 
(estimated bandwidth $<$ current representation bitrate)\big)\Big)} 
\State Switch down
\ElsIf{
                \Big(next representation bitrate $<$ estimated bandwidth) and \newline 
\phantom{aaaaa} \Big(\big( buffer level $\geq$ almost full \big) or \newline
 \phantom{aaaaaaa}    \big(buffer level is safe and is filling\big)\Big)} 
\State Switch up	
\EndIf
\EndIf
\end{algorithmic}
\end{algorithm}


\subsection{Geo-MaxBW Adaptation Logic}
\label{Geo-MaxBW Adaptation Logic}

The MaxBW algorithm \cite{KLUDCP} adaptive decision is based on the
measured download time of each segment and the average measured
bitrate of the whole session. The Geo-MaxBW algorithm uses a crowd-based
bandwidth estimate as presented in Algorithm \ref{alg:algo0}.

\subsection{Geo-Predictive Adaptation Logic (GPAL)}
\label{Geo-Predictive Adaptation Logic (GPAL)}

The GPAL algorithm, Algorithm \ref{Algorithm:GPAL}, determines the representation
of the next media segment to be fetched. The algorithm estimates the
current segment download path based on the client's location and
speed. It predicts the future path network bandwidth conditions based
on the client's playout buffer and the crowd estimated throughput. The
algorithm calculates the playout buffer fullness ratio ($B_{p}$) based
on the maximum between the current buffer levels divided by the
maximum buffer size allocation and 10\%. We used the 10\% to
select a higher bandwidth when the playout buffer is drained.

\begin{algorithm}
\caption{GPAL: Geo Predictive Adaptation Logic Algorithm} 
\label{Algorithm:GPAL}
\begin{algorithmic}[1]
\State \textsl{$\rho$}: predicted mobile bandwidth for next segment.
\State \textsl{$B(t)$}: current playout buffer duration.
\State \textsl{$B_{p}$}: playout buffer fullness ratio. 
\State \textsl{$B_{max}$}: maximum buffer size.
\State \textsl{$\tau$}: selected quality for download.
\State $B_{p} =  \frac{B(t)}{B_{max}}$
\If{first segment}
\State $B_{p}$ = 0.5
\EndIf
\State \textsl{Estimate $X_{bw}(t)$ for each segment download.}
\State \textsl{$\rho$} = $X_{bw}(t)\cdot B_{p}$
\State $\tau$ = the highest bit rate representation for which $\tau < \rho$
\If {$B_{p}$ = 0.2}
\State Reduce $\tau$ in one representation.
\EndIf
\State return $\tau$
\end{algorithmic}
\end{algorithm}

\section{Dataset}
\label{datasets}

The WeFi dataset contains $336,551$ samples from the California I110
and I405 interstates. The I110 is an interstate highway in the Los
Angeles area and connects San Pedro and the port of Los Angeles with
downtown Los Angeles and Pasadena. The I405 is a major north-south
interstate highway in Southern
California. Table \ref{tab:InterstateSummery} summarizes the
interstates' general features.
\begin{table}[h!]
  \centering
  \begin{tabular}{| l | l | l |}\hline	
									 &I110 				& I405 \\ \hline
	Date of creation & $1-7.12.2014$&	$1-7.12.2014$	 \\ \hline
	Number of samples& $125079$			& $211472$	 \\ \hline
	Section length   & $30$	$km$		& $17$ $km$	  \\ \hline
	Number of users  & $5838$				& $6170$      \\ \hline
	Number of samples& $125079$			& $211472$    \\ \hline
  \end{tabular}
  \caption{Interstate roads summary}
  \label{tab:InterstateSummery}
\end{table}

A large number of different applications generated the data.
Most of the applications either regularly send low rate updates or are
in the idle state (sending keep-alive messages).

We estimate the average throughput (bits per second) per sample $s$
for the interval $x$ using Eq \ref{eq:normalization}.

\begin{figure}[htbp]
	\centering
		\includegraphics[scale =0.35]{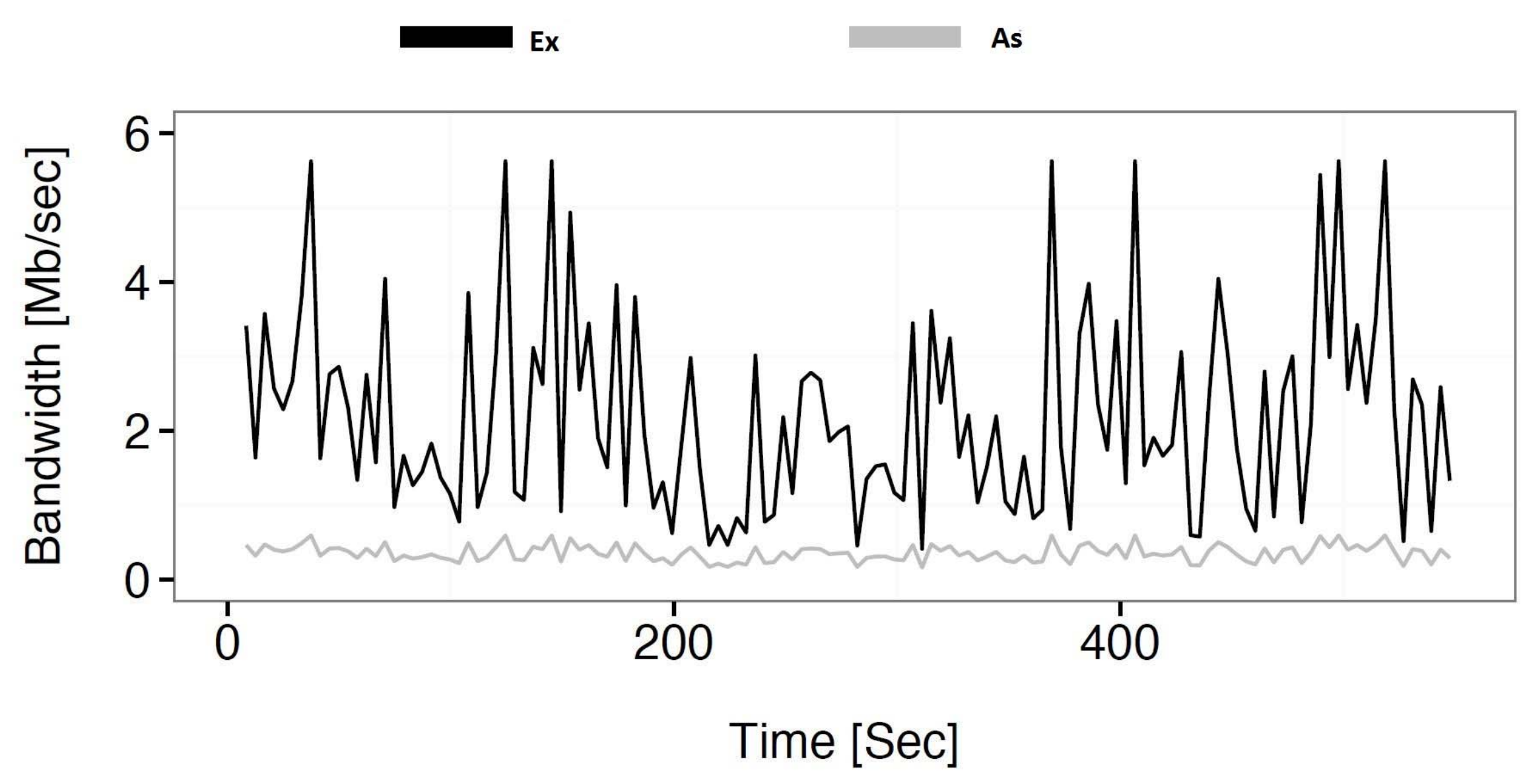}
	\caption{I110 average throughput per sample for an interval ($E_x$)
vs. the throughput ($A_s$)}
	\label{fig:i110_Before_After_Analysis}
\end{figure}

\begin{figure}[htbp]
	\centering
		\includegraphics[scale =0.35]{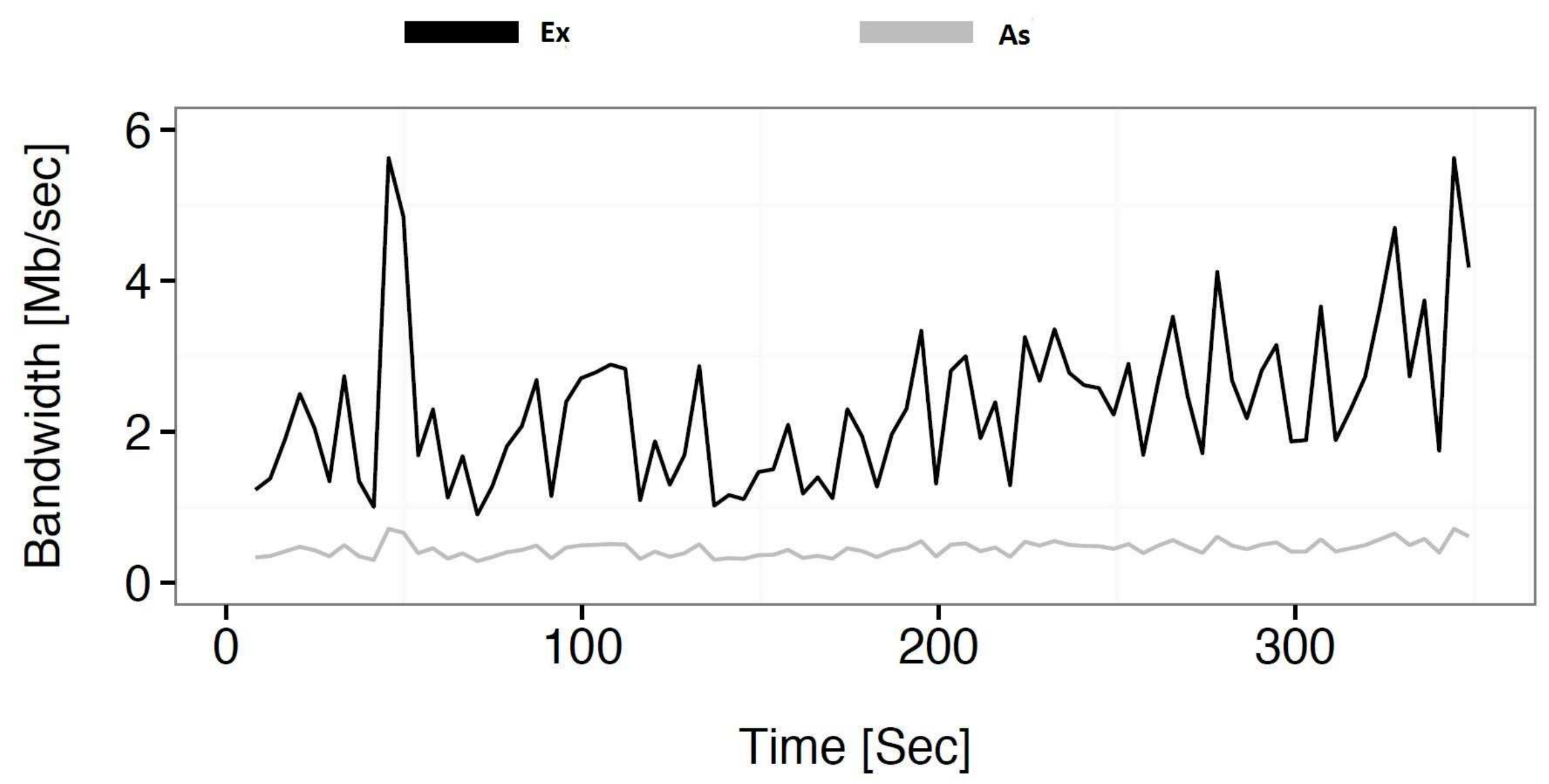}
	\caption{I405 average throughput per sample for an interval ($E_x$)
vs. the throughput ($A_s$)}
	\label{fig:i405_Before_After_Analysis}
\end{figure}

\begin{center}
\begin{equation}
\label{eq:normalization}
\begin{split}
E_{x} = \frac{\sum_{s \in x}D_{s} \cdot A_{s}}{\sum_{s \in x}D_{s}}\\
\end{split}
\end{equation}
\end{center}
where $D_{s}$ is the total data received in sample $s$ and $A_{s}$ is
the average throughput in sample $s$.  Figs
\ref{fig:i110_Before_After_Analysis}-\ref{fig:i405_Before_After_Analysis}
illustrate the average throughput per sample for an interval ($E_x$)
vs. the throughput ($A_s$). In these figures, each path is divided into
$12$ meter segments.


\subsection{Interstate I110}
\label{Interstate I110}

\begin{figure*}[htbp]
\centering
\subfigure[I110 throughput heat map.]{\label{fig:i110heatmap}\includegraphics[width=0.3\textwidth,height= 0.2\textwidth]{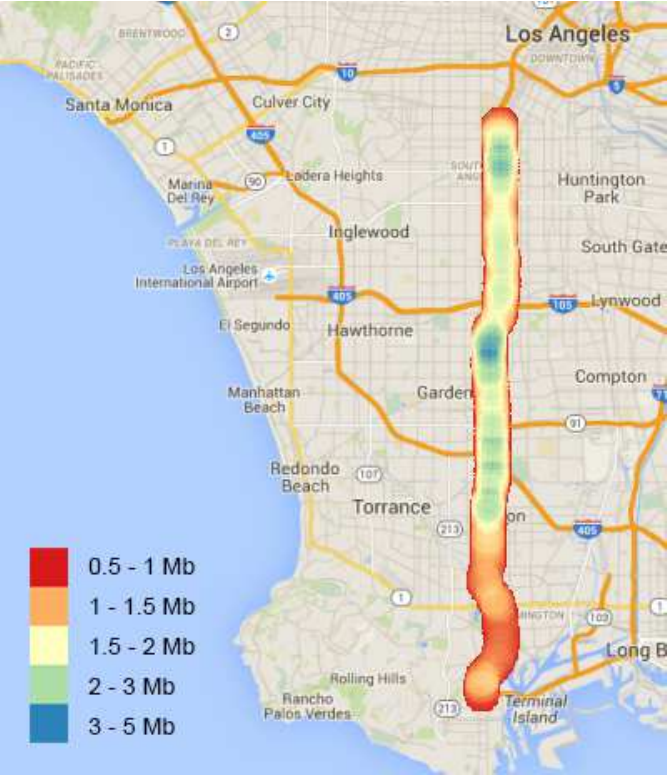}}
\subfigure[I110 path positioning]{\label{fig:i110pathpositioning}\includegraphics[width=0.5\textwidth]{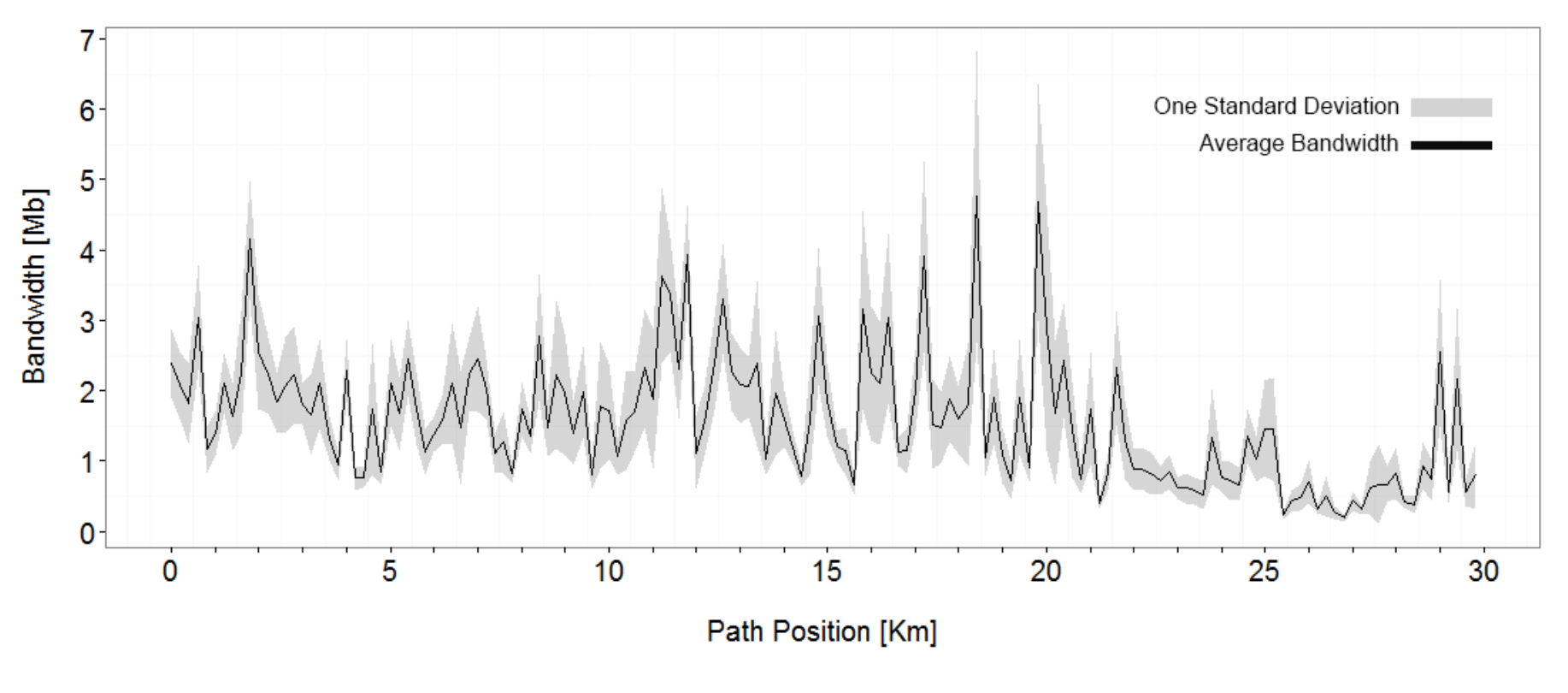}}
\subfigure[Bandwidth entropy (PMF) analysis]{\label{fig:i110BandwidthEntropyAnalyze}\includegraphics[width=0.3\textwidth]{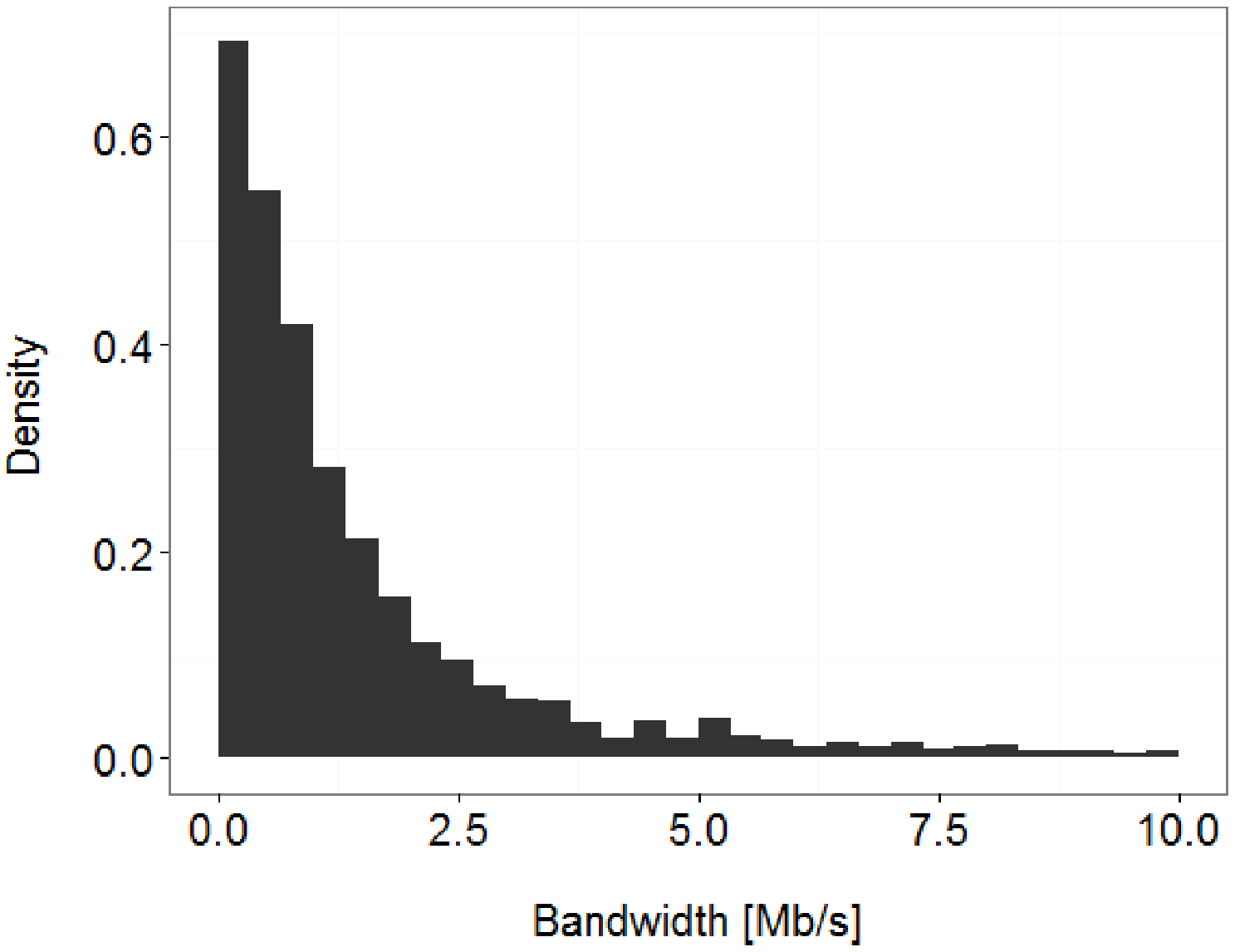}}
\subfigure[Samples density entropy (PMF) analysis]{\label{fig:i110dansityEntropyAnalyze}\includegraphics[width=0.3\textwidth]{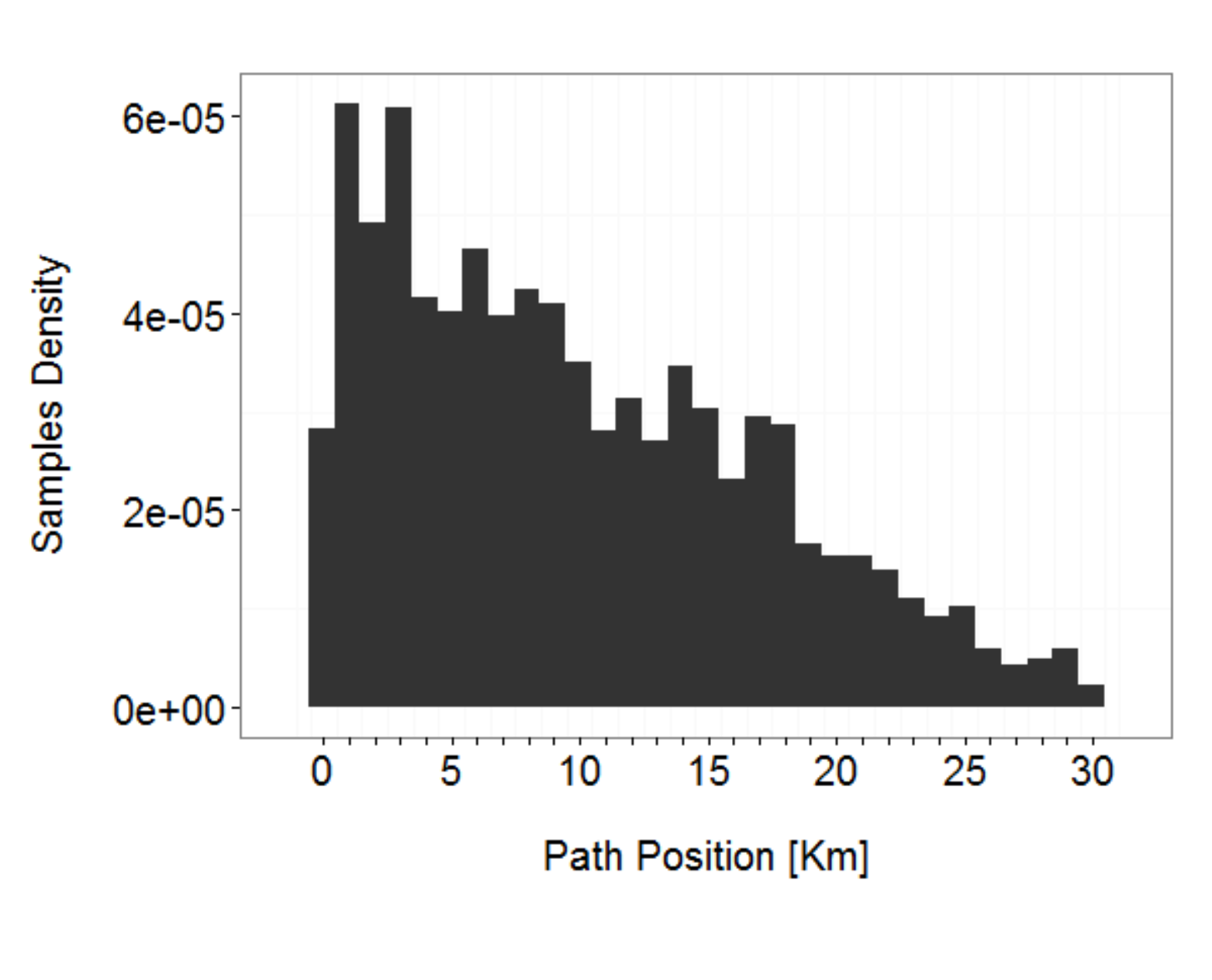}}
\subfigure[Route average throughput between 3:00 - 9:00]{\label{fig:i110throughput0309}\includegraphics[width=0.3\textwidth]{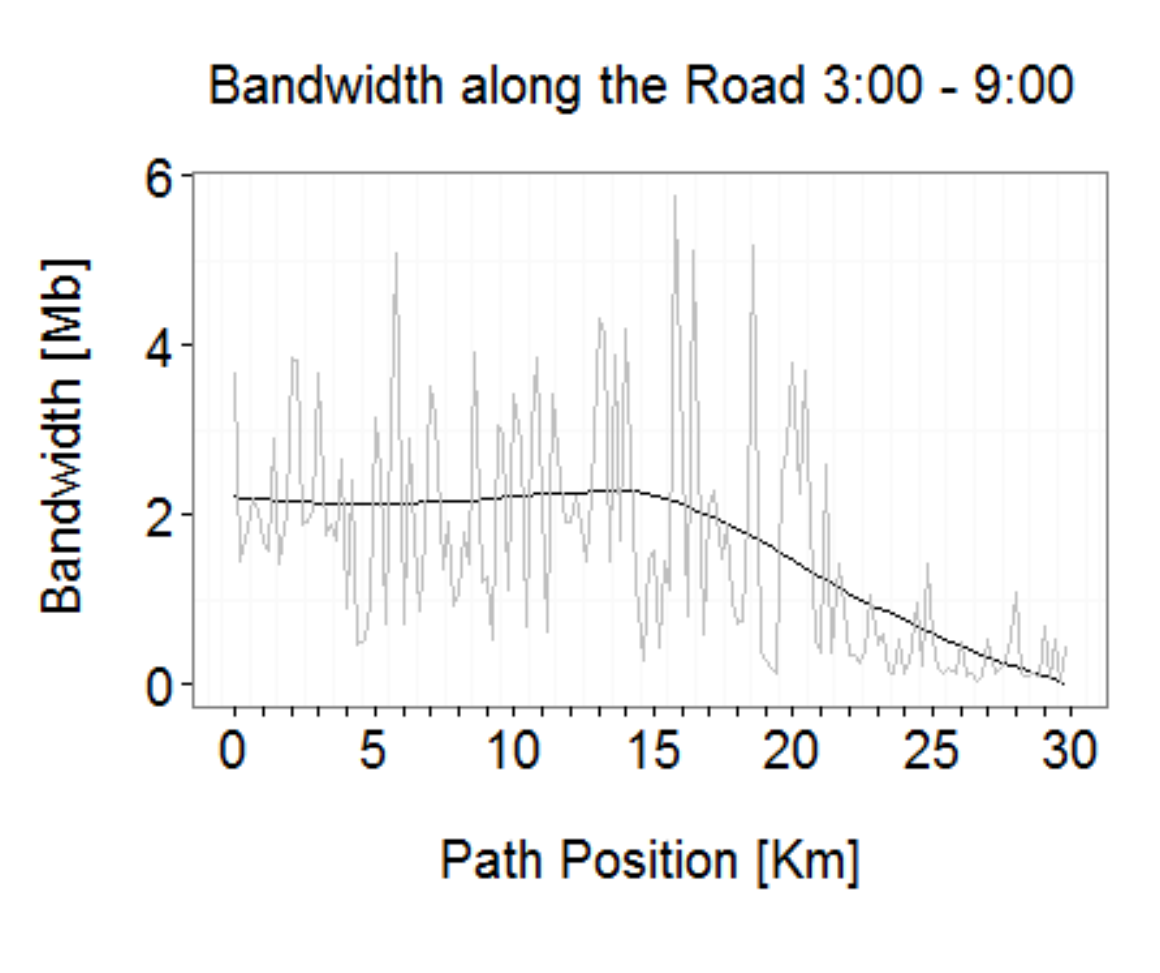}}
\subfigure[Route average throughput between 9:00 - 15:00]{\label{fig:i110throughput0915}\includegraphics[width=0.3\textwidth]{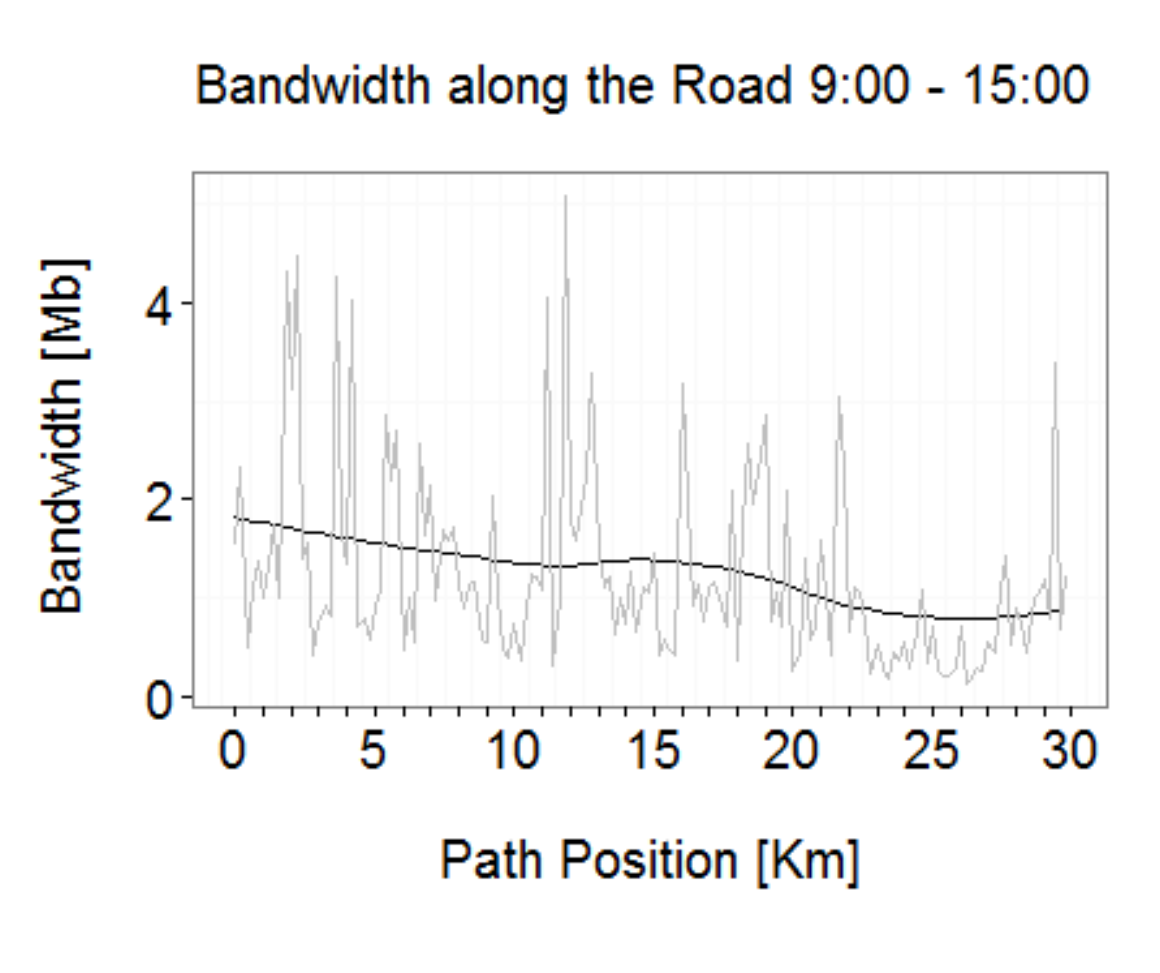}}
\subfigure[Route average throughput between 15:00 - 21:00 ]{\label{fig:i110throughput1521}\includegraphics[width=0.3\textwidth]{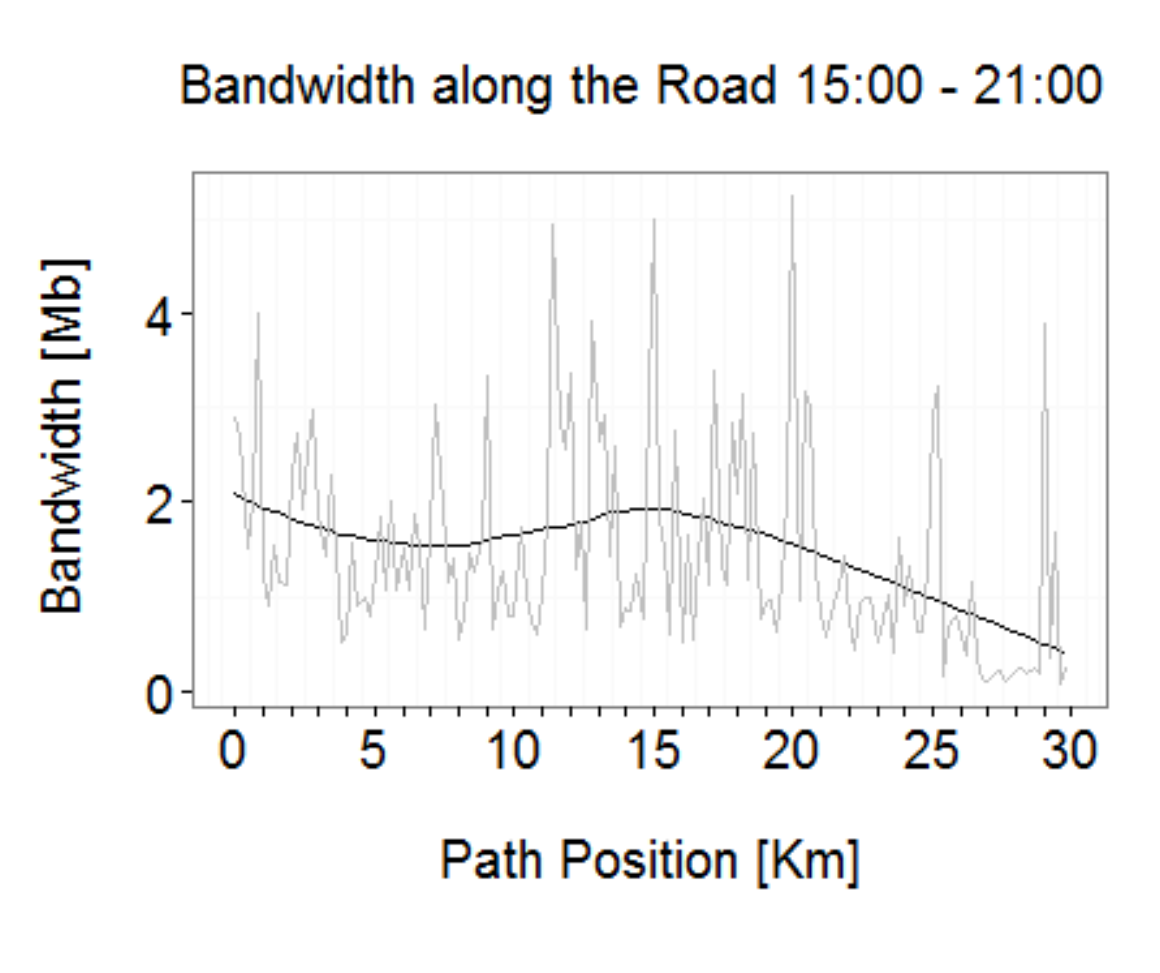}}
\subfigure[Route average throughput between 21:00 - 03:00 ]{\label{fig:i110throughput2130}\includegraphics[width=0.3\textwidth]{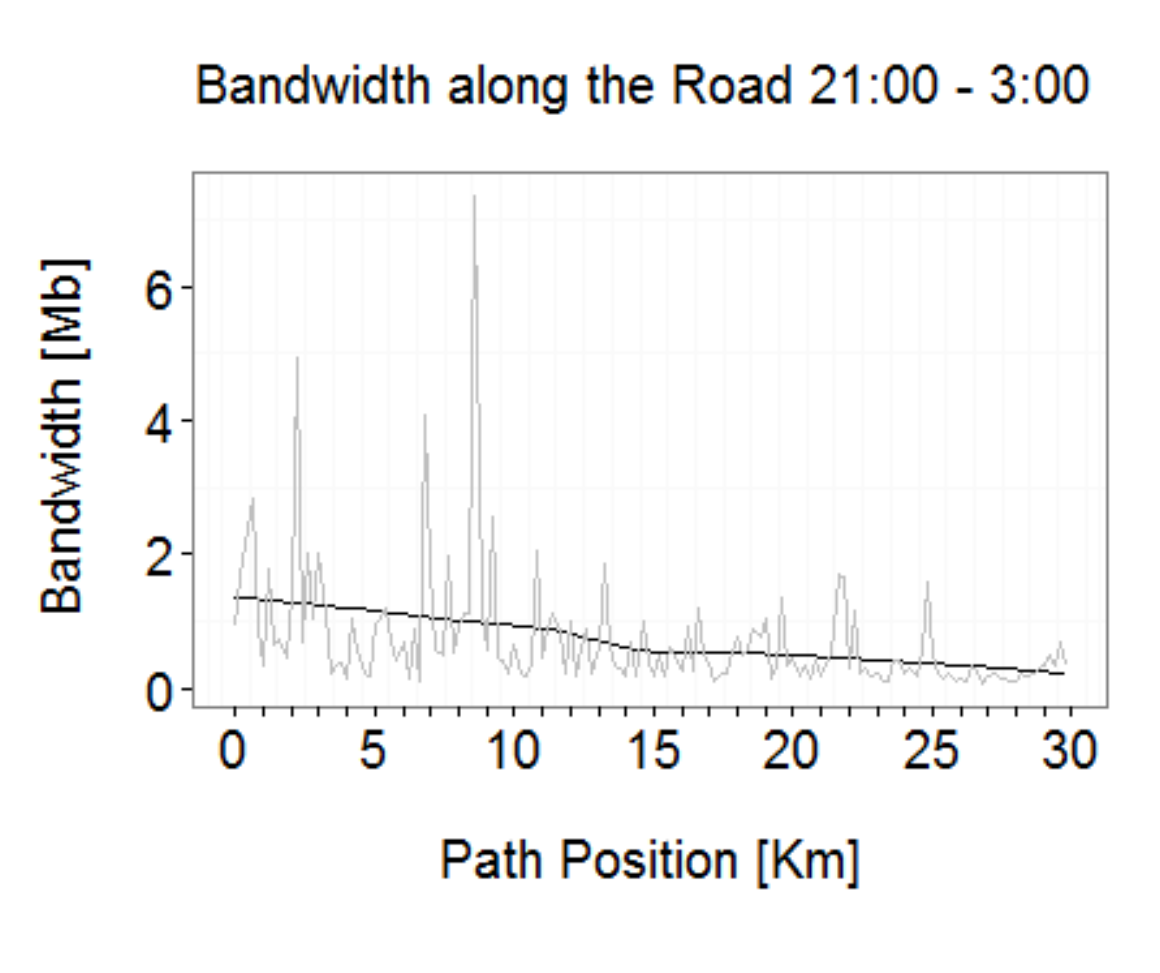}}
  \caption{Interstate I110 dataset detailed depiction.}
   \label{fig:I110Info}
\end{figure*}

The interstate heat map is illustrated in Fig. \ref{fig:i110heatmap}
which shows that the road throughput can vary between $0.5 - 5 \lbrack
Mb/s\rbrack$. Fig. \ref{fig:i110pathpositioning} depicts the measured
bandwidth of the path (average and STD). We define this bandwidth path
as I110.  

The median throughput of the interstate is $0.86 \lbrack Mb/s\rbrack$,
the average throughput is $1.585 \lbrack Mb/s\rbrack$ and the STD is
$2$. That is, the path has many fluctuations. Thus, it is challenging
for adaptive streaming clients to adapt to its network conditions.

Fig. \ref{fig:i110BandwidthEntropyAnalyze} depicts the throughput
density and the sample densities along the route. We split the
throughput density into fixed bins from $0$ to the maximum observed
throughput, $10 \lbrack Mb/s\rbrack$. It is clear that lower
throughput in the ranges of $0 - 2 \lbrack Mb/s\rbrack$ are more
likely while throughput above $5 \lbrack Mb/s\rbrack$ are less
common.  Fig. \ref{fig:i110dansityEntropyAnalyze} shows the sample
densities along the route. From $23 km$ the sample densities
decrease. Figs. \ref{fig:i110throughput0309} -
\ref{fig:i110throughput2130} show the throughput behavior in different
time ranges. Obviously, the demand for
bandwidth at night (21:00 - 3:00) is much lower than during rush
hour. Table. \ref{tab:InterstateI110AverageBitRateInDifferentHours}
summarizes the average bit rate at different time periods.

\begin{figure*}[htbp]
\centering
\subfigure[I405 throughput heat map.]{\label{fig:i405heatmap}\includegraphics[width=0.3\textwidth,height= 0.2\textwidth]{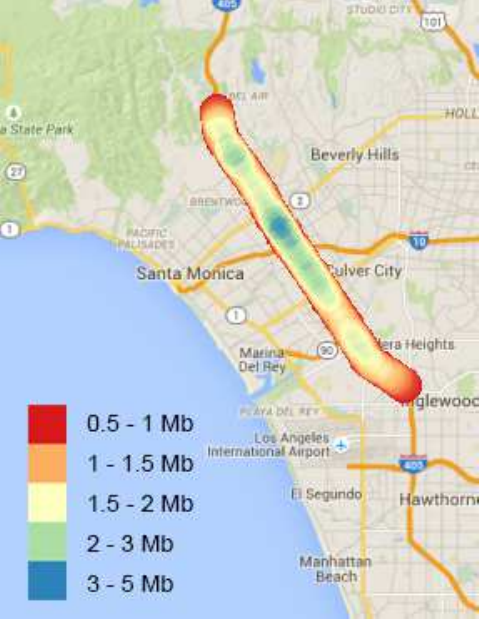}}
\subfigure[I405 path positioning]{\label{fig:i405pathpositioning}\includegraphics[width=0.5\textwidth]{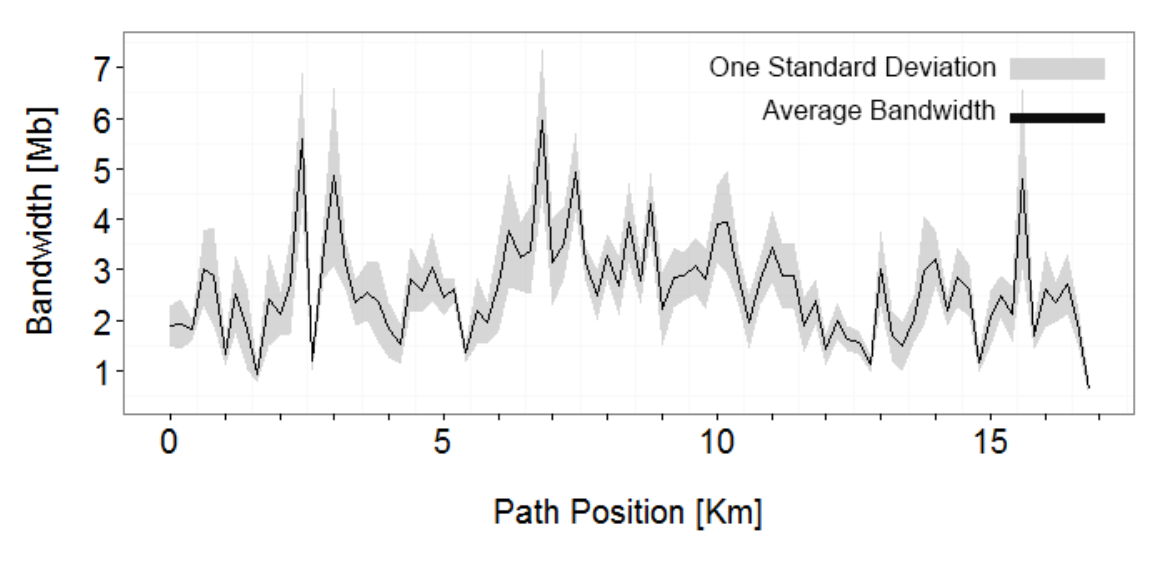}}
\subfigure[Bandwidth entropy (PMF) analysis]{\label{fig:i405BandwidthEntropyAnalyze}\includegraphics[width=0.3\textwidth]{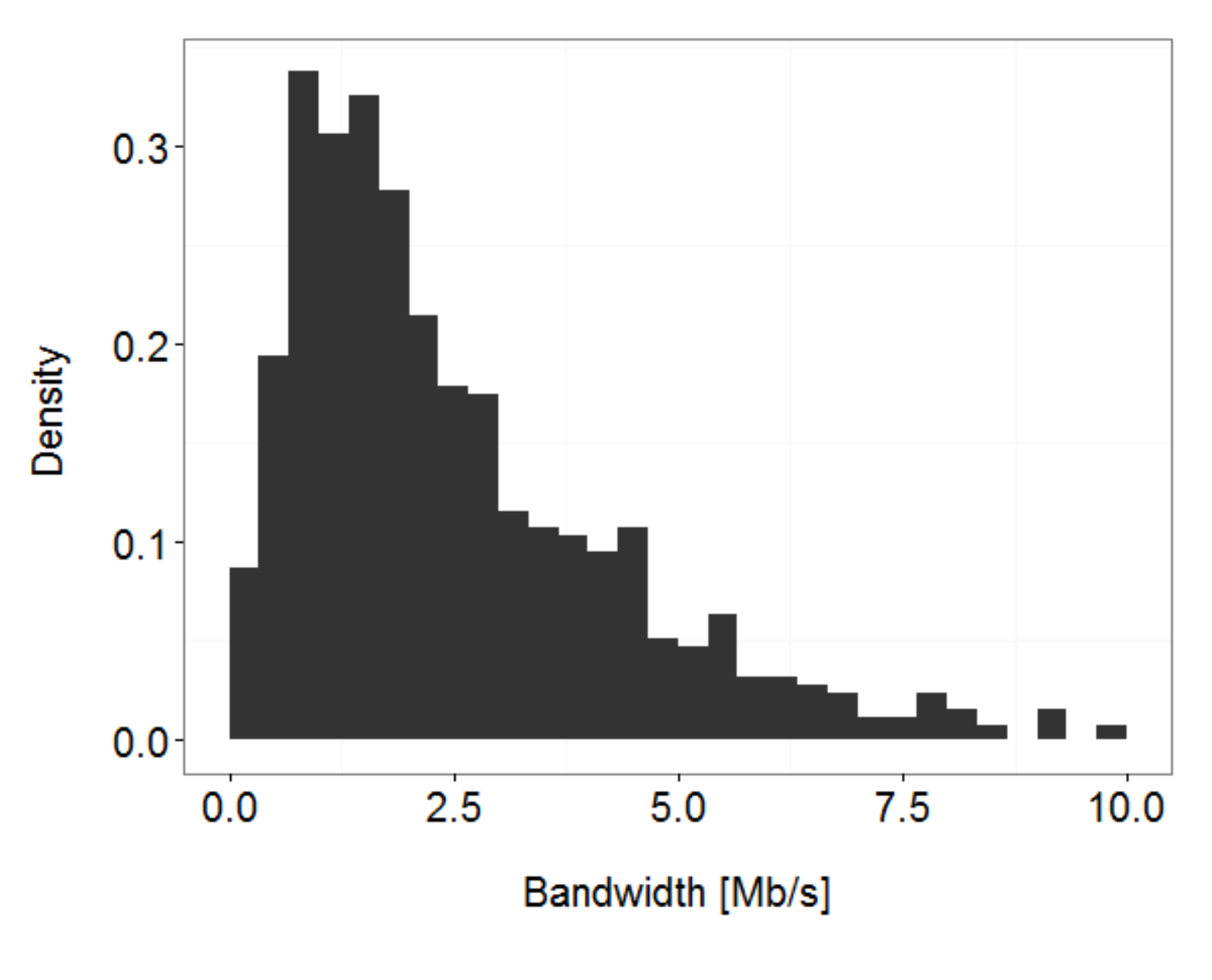}}
\subfigure[Samples density entropy (PMF) analysis]{\label{fig:i405dansityEntropyAnalyze}\includegraphics[width=0.3\textwidth]{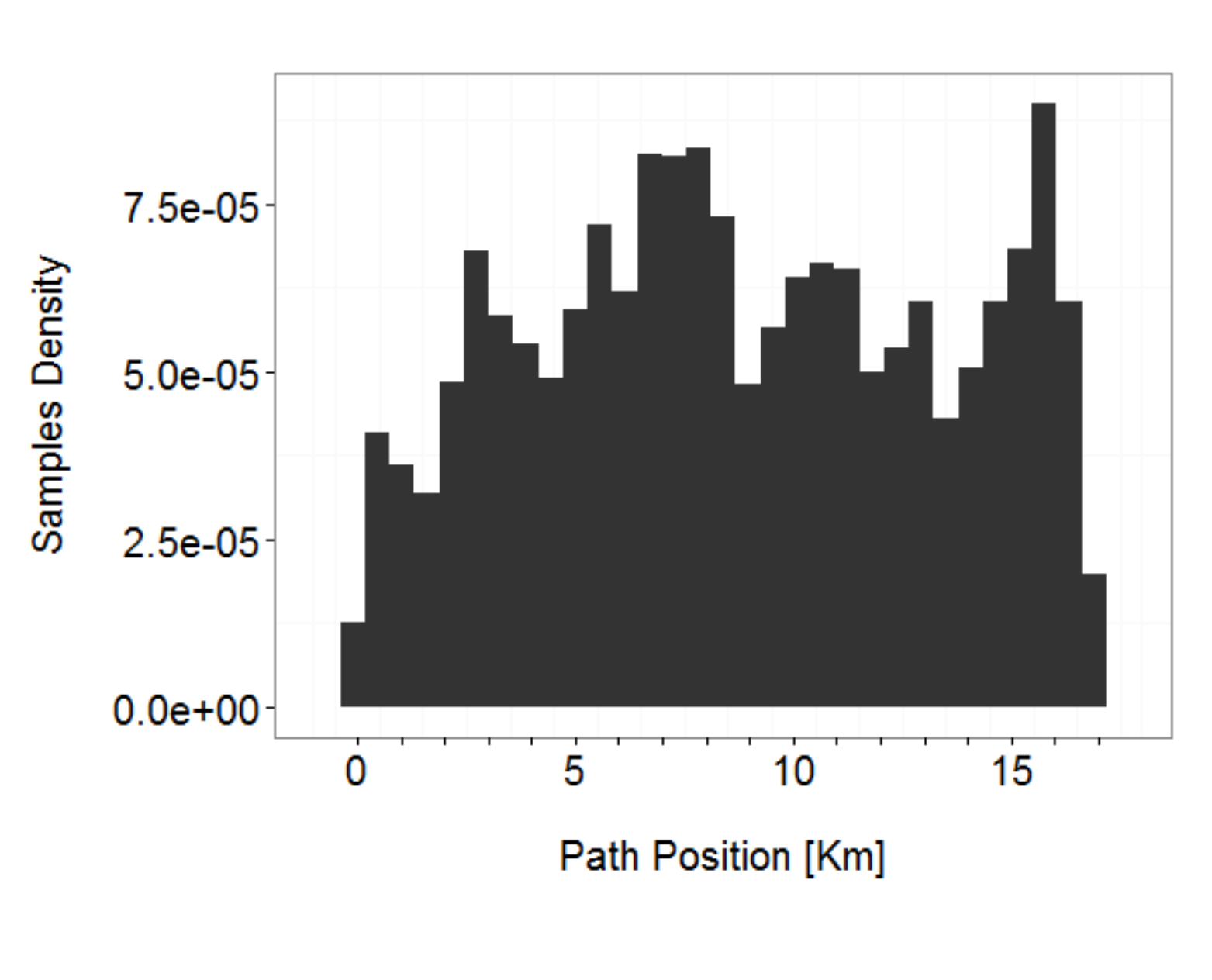}}
\subfigure[Route's throughput average between 3:00 - 9:00]{\label{fig:i405throughput0309}\includegraphics[width=0.3\textwidth]{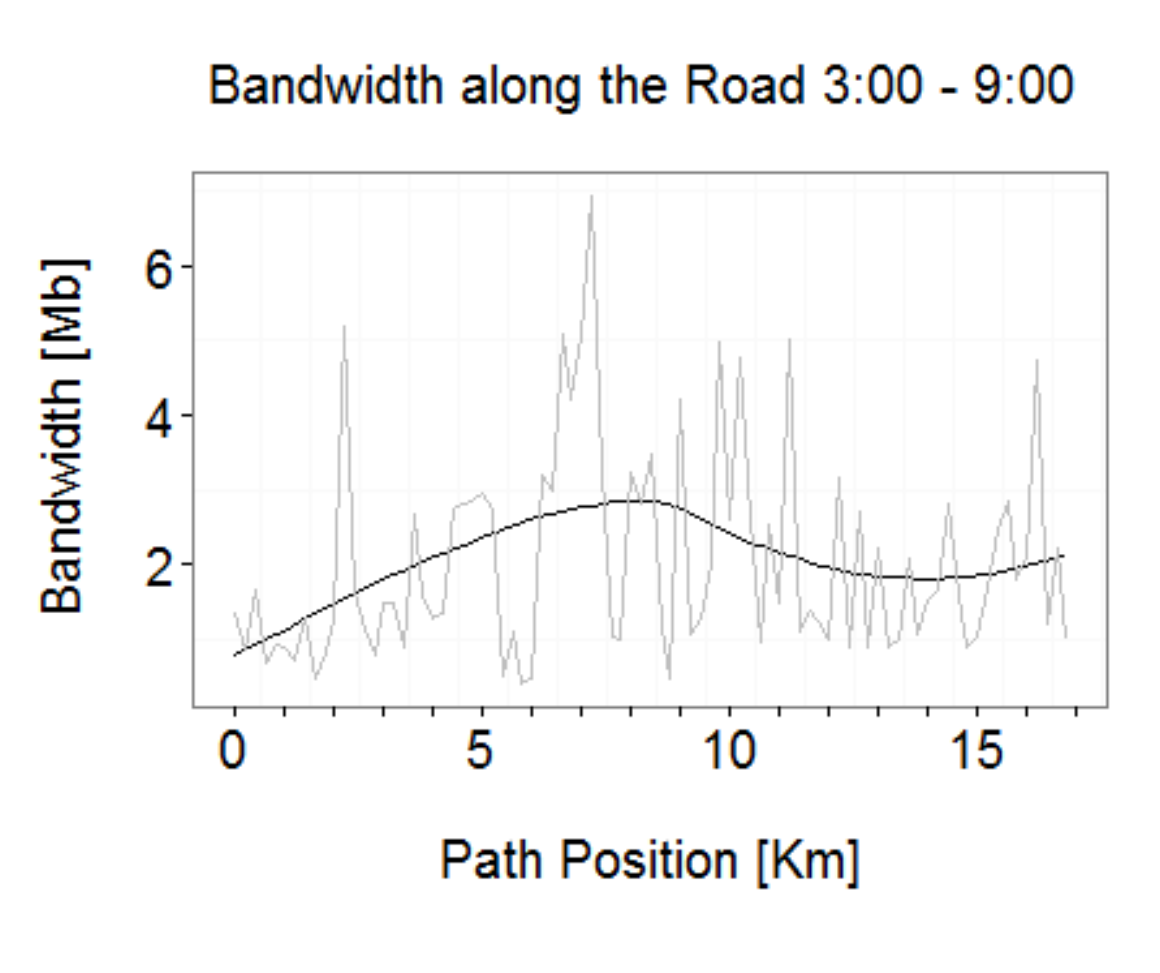}}
\subfigure[Route's throughput average between 9:00 - 15:00]{\label{fig:i405throughput0915}\includegraphics[width=0.3\textwidth]{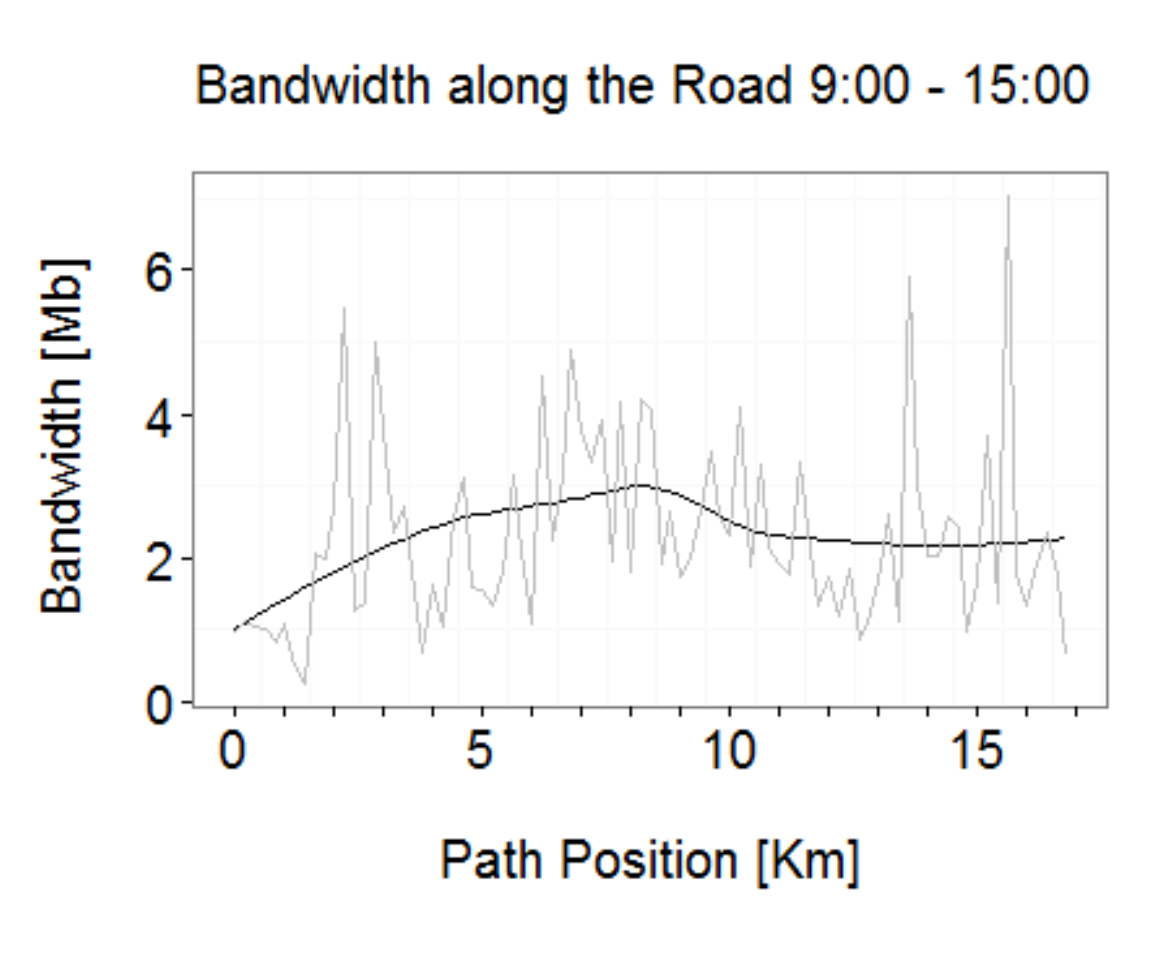}}
\subfigure[Route's throughput average between 15:00 - 21:00 ]{\label{fig:i405throughput1521}\includegraphics[width=0.3\textwidth]{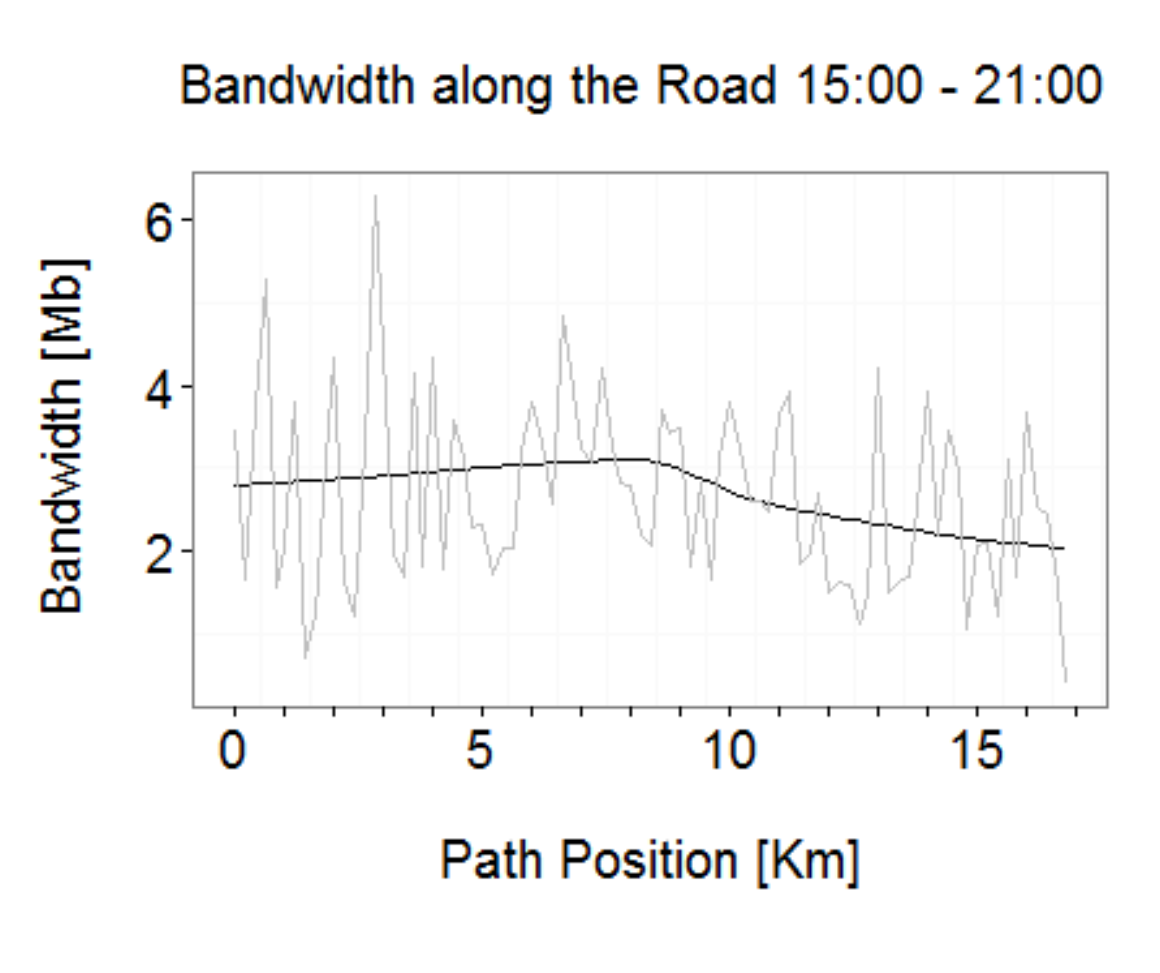}}
\subfigure[Route's throughput average between 21:00 - 3:00 ]{\label{fig:i405throughput2103}\includegraphics[width=0.3\textwidth]{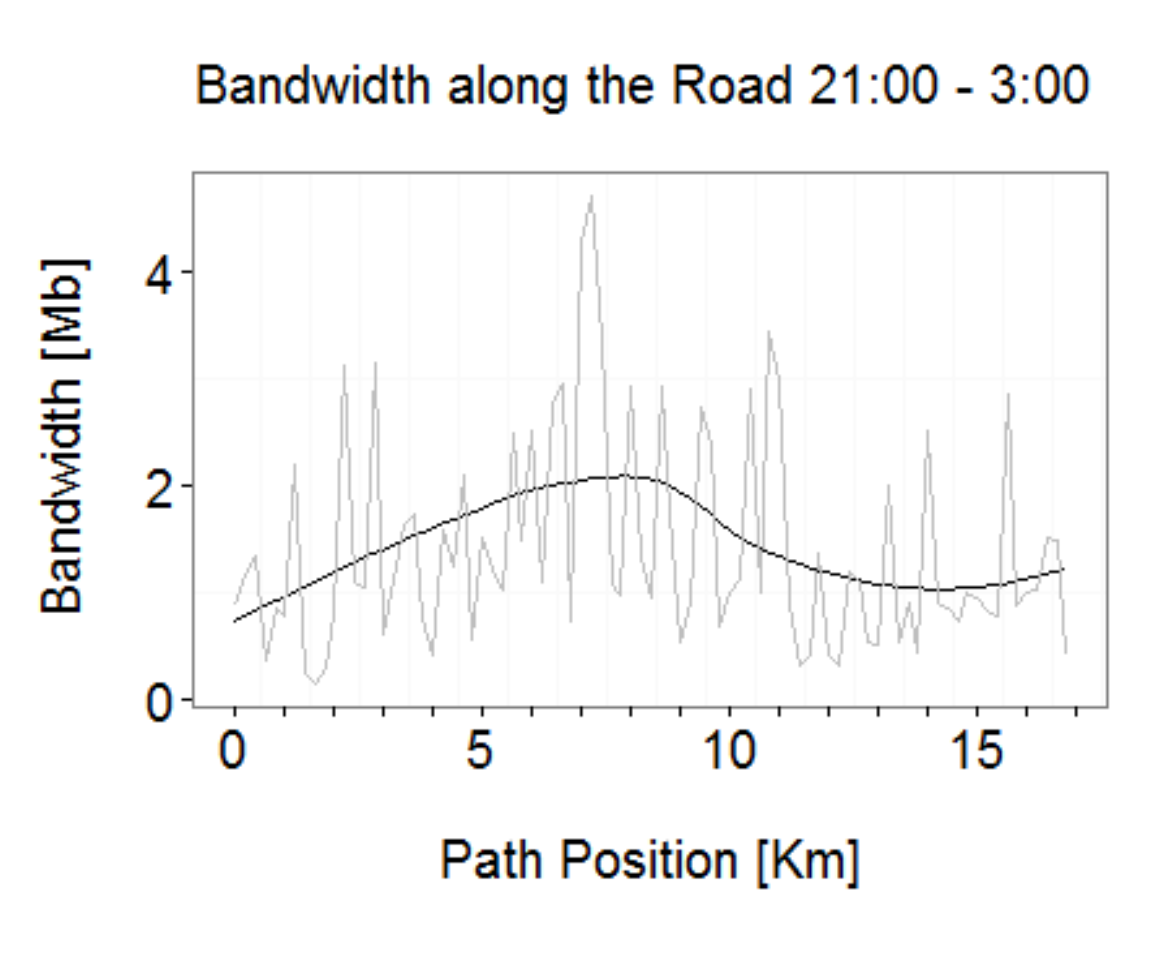}}\caption{Interstate I405 dataset detailed depiction.}
\label{fig:i405Info}
\end{figure*}

\begin{table}[htbp]
  \centering
  \begin{tabular}{| l | l |}\hline	
	Time range & Average bit rate \lbrack Mb/s\rbrack \\ \hline
	3:00-9:00 & $1.6$\\\hline
	9:00-15:00 & $1.24$\\\hline
	15:00-21:00 & $1.46$\\\hline
	21:00-3:00 & $0.72$\\\hline
  \end{tabular}
  \caption{Interstate I110 average bit rate at different hours}
  \label{tab:InterstateI110AverageBitRateInDifferentHours}
\end{table}

\subsection{Interstate I405}
\label{Interstate I405}

The I405 interstate is shorter but has a higher number of samples than
the I110 interstate (see Table \ref{tab:InterstateSummery}).  The
interstate heat map is illustrated in Fig. \ref{fig:i405heatmap} which
shows that the road throughput varies between $0.5 - 5 \lbrack
Mb/s\rbrack$. Fig. \ref{fig:i405pathpositioning} depicts the measured
bandwidth of the path (average and STD). We define this bandwidth path
as $I405A$.

The median throughput of the interstate is $1.97 \lbrack Mb/s\rbrack$,
the average throughput is $2.63 \lbrack Mb/s\rbrack$ and the STD is
$2.15$. The I405 interstate has a higher throughput average than
I110. The STD is slightly higher.

Fig. \ref{fig:i405BandwidthEntropyAnalyze} illustrates the throughput
density and the sample densities along the route. We split the
throughput density into fixed bins from $0$ to the maximum observed
throughput $10 \lbrack Mb/s\rbrack$. The table shows that the I405
throughput density is different from the I110 throughput density and
the throughput is better spread between $0.5-2.5 \lbrack Mb/s\rbrack$.
Fig. \ref{fig:i405dansityEntropyAnalyze} depicts the density of the
samples along the route. This road is more evenly dense than
I110. Figs. \ref{fig:i405throughput0309} -
\ref{fig:i405throughput2103} show the throughput behavior at different
time periods. It shows that the throughput demand on this road is
higher even in the late hours (Fig. \ref{fig:i405throughput2103}).

\section{Experiments and Results}
\label{Experiments and Results}

We describe our experimental setup and video representation
information in Section \ref{Setup and Evaluation}. We discuss our
experimental results Sections \ref{Experimental Results}, \ref{Interstate
  I110 Results} and \ref{Interstate I405 Results}.

\subsection{Experimental Setup}
\label{Setup and Evaluation}

This section describes our experimental settings and video encoding
configuration. We used the Big Buck Bunny (BBB) \cite{BBB} video encoded
into fixed duration segments of $2$ seconds. Table \ref{tab:video
  representation} illustrates the BBB available representation stored in
the streaming server. The client playout buffer duration was set to
$30$ seconds.

\begin{table*}[htbp]
  \centering
  \begin{tabular}{| l | l | l | l |}\hline
    Representation& SSIM & PSNR $\lbrack dB\rbrack$ & Average bit rate $\lbrack Kb/s\rbrack$ \\\hline
	$50$   & $0.719$ & $24.4$   & $51.05$   \\\hline
	$100$  & $0.8$   & $28.3$   & $98.91$   \\\hline
	$200$  & $0.89$	 & $32.4$   & $193.31$  \\\hline
	$250$  & $0.914$ & $34$     & $240.96$  \\\hline
	$500$  & $0.96$  & $38$     & $480.15$  \\\hline
	$750$  & $0.971$ & $40$     & $721.56$  \\\hline
	$1000$ & $0.977$ & $41.4$   & $964.16$  \\\hline
	$1500$ & $0.985$ & $0.91$   & $1452.44$ \\\hline
	$2000$ & $0.988$ & $44.5$   & $1942.4$  \\\hline
	$2400$ & $0.989$ & $45.28$  & $2335.2041$  \\\hline
  \end{tabular}
  \caption{Big Buck Bunny representation information}
  \label{tab:video representation}
\end{table*}	

\begin{figure}[h!]
  \centering
  \includegraphics[width=0.5\textwidth]{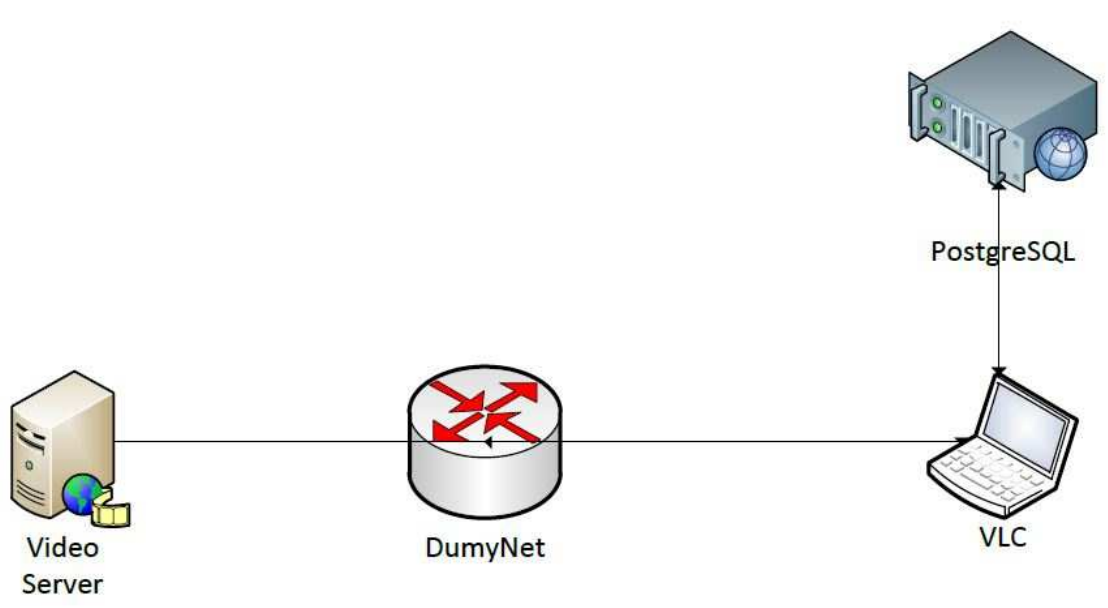}
  \caption{Experimental setup diagram}
  \label{fig:geoPredictive}
\end{figure}

Fig. \ref{fig:geoPredictive} illustrates our experimental
setup. First, the user requests (VLC \cite{VLC_22}) the video MPD file
from the HTTP server. After the client receives it, the adaptation
logic algorithm requests the crowd estimate from the PostgreSQL
geo-predictive server. Then, the user sends a request to the server
using a simple API implementation which only sends the following
information to the server: the search
radius ($250$ meters), the user's current location and the estimated end
point (which depends on the user's average speed). The geo-predictive
module predicts the average throughput. Since this API is very
lightweight, the process delay is negligible. We do not assume we know
the route. Therefore, we used a batch fetching mechanism. That is,
before the current segment download ends, we fetch the crowd estimate
for the next segment. Each adaptation logic can analyze the data or
use the API differently but the fetching optimization is beyond
the scope of this study.  The DumyNet \cite{Dum_24} shapes the traffic
according to the network scenario. As a result, the segment download
is delayed according to the network conditions. In order to compare
our work to state-of-the-art algorithms we used the same segment fetching
schema as these works, where the client downloads each segment one
after the other.

\subsection{Experimental Results}
\label{Experimental Results}
	
In Eq. \ref{eq:Qualitymaximization2} we stated our goal. In our
experiments $B_{max}$ was 30 seconds. All compared algorithms realized
the constraints of Eq. \ref{eq:Qualitymaximization2} (as can be seen
in
Figs. \ref{fig:I110AtestResultComparison}-\ref{fig:I405AtestResultComparison}). Table
\ref{tab:Average_eMOS} summarizes the average eMOS score
\cite{claeys2014design} for all the algorithms. The table shows that
our GPAL algorithm outperforms all other algorithms. Additionally, the
integration of crowd information into state-of-the-art algorithms
boosts their performance. Geo-MaxBW eMOS score was $4.35$ whereas the
non-crowd-based MaxBW eMOS score was only $4.21$. Similarly,
Geo-MAL eMOS score was $3.74$ whereas the non-crowd-based MAL eMOS
score was only $3.41$.

\begin{table}[t!]
  \centering
  \begin{tabular}{| l | l |}\hline
  Algorithm & Avergae eMOS  \\\hline
	GPAL (\textbf{our}, Section \ref{Geo-Predictive Adaptation Logic (GPAL)})                  & $4.39$ \\\hline
	MaxBW (\cite{KLUDCP})                                & $4.21$ \\\hline
	Geo-MaxBW (\textbf{our} adaptation of \cite{KLUDCP}, Section \ref{Geo-MaxBW Adaptation Logic}) & $4.35$\\\hline
	MAL (\cite{KLUDCP})                                  & $3.41$ \\\hline
	Geo-MAL (\textbf{our} adaptation of \cite{MAL_conf}, Section \ref{Geo-MAL}) & $3.74$ \\\hline
	1-Predict (\cite{hao2014gtube})                      & $3.24$ \\\hline
	n-Predict (\cite{hao2014gtube})                      & $2.15$ \\\hline
	PBA (\cite{Zou2015})                                 & $2.89$\\\hline
	MASERATI (\cite{han2013maserati})                    & $1.37$ \\\hline
  \end{tabular}
  \caption{Average eMOS score for all algorithms}
  \label{tab:Average_eMOS}
\end{table}

\subsection{Detailed Results of Interstate I110 and Discussion}
\label{Interstate I110 Results}

\begin{figure*}
\centering
\subfigure[GPAL]{\label{fig:i110GPALbandwidthEtsimation}\includegraphics[width=0.3\textwidth]{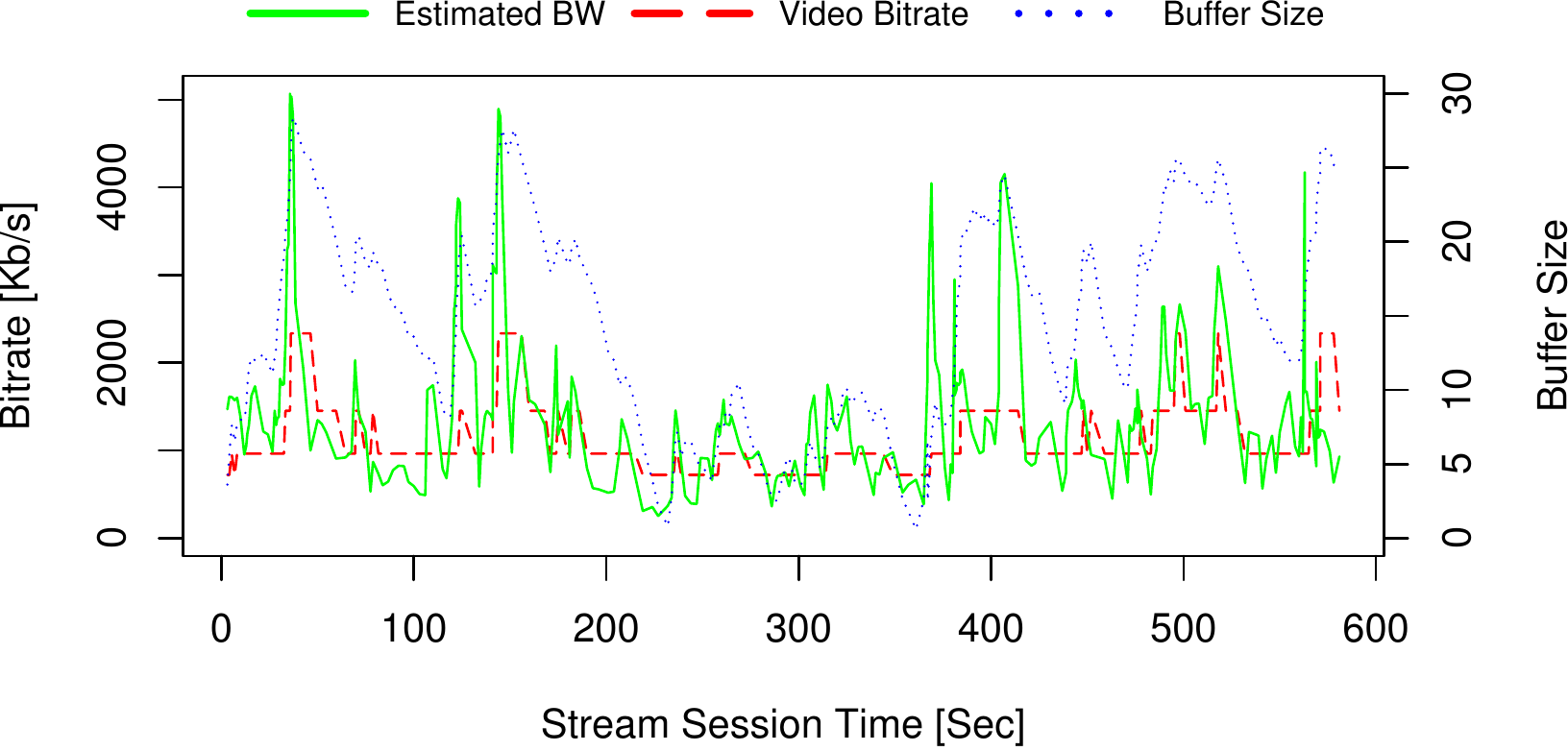}}
\subfigure[MaxBW]{\label{fig:i110MaxBwbandwidthEtsimation}\includegraphics[width=0.3\textwidth]{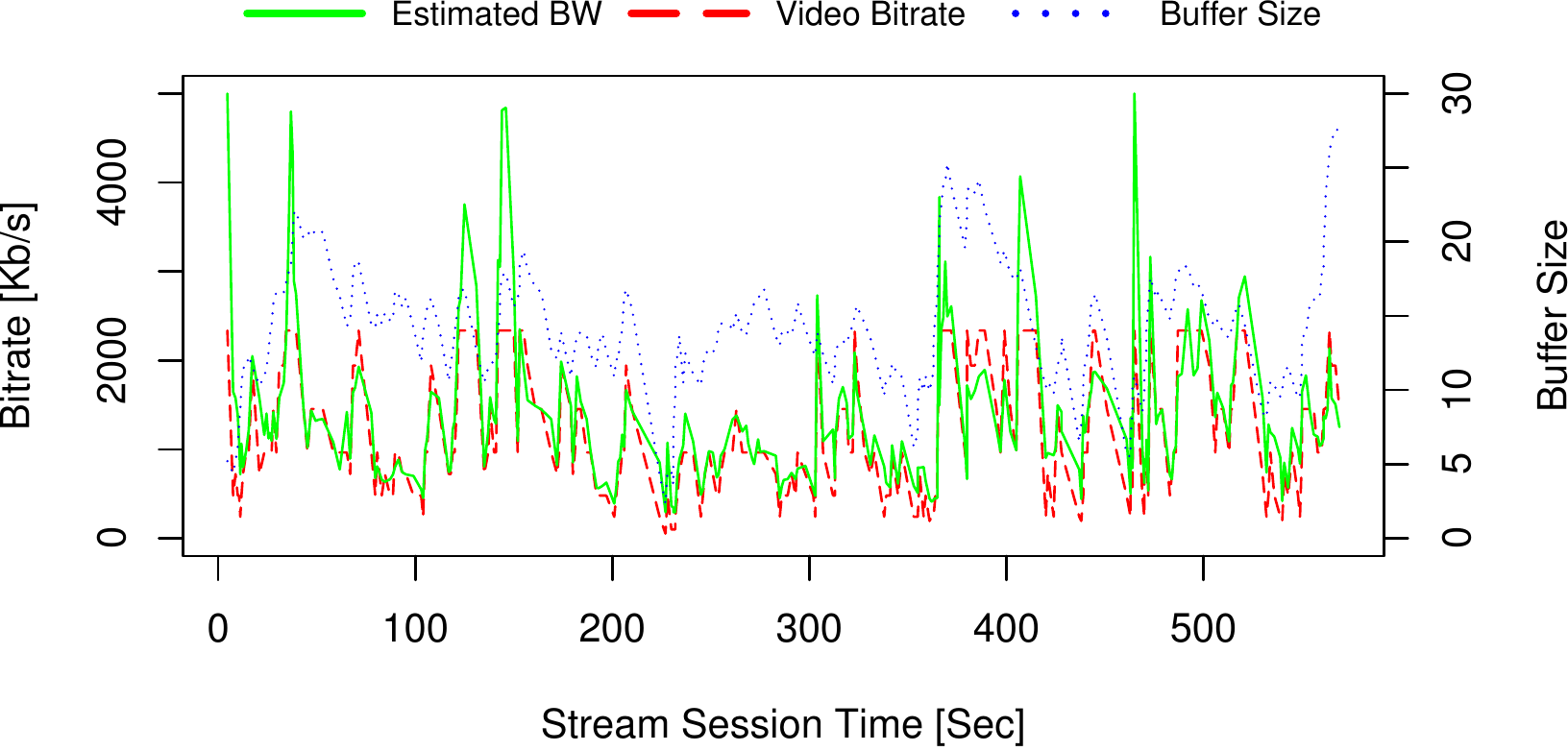}}
\subfigure[Geo-MaxBW]{\label{fig:i110GeoMaxBWbandwidthEtsimation}\includegraphics[width=0.3\textwidth]{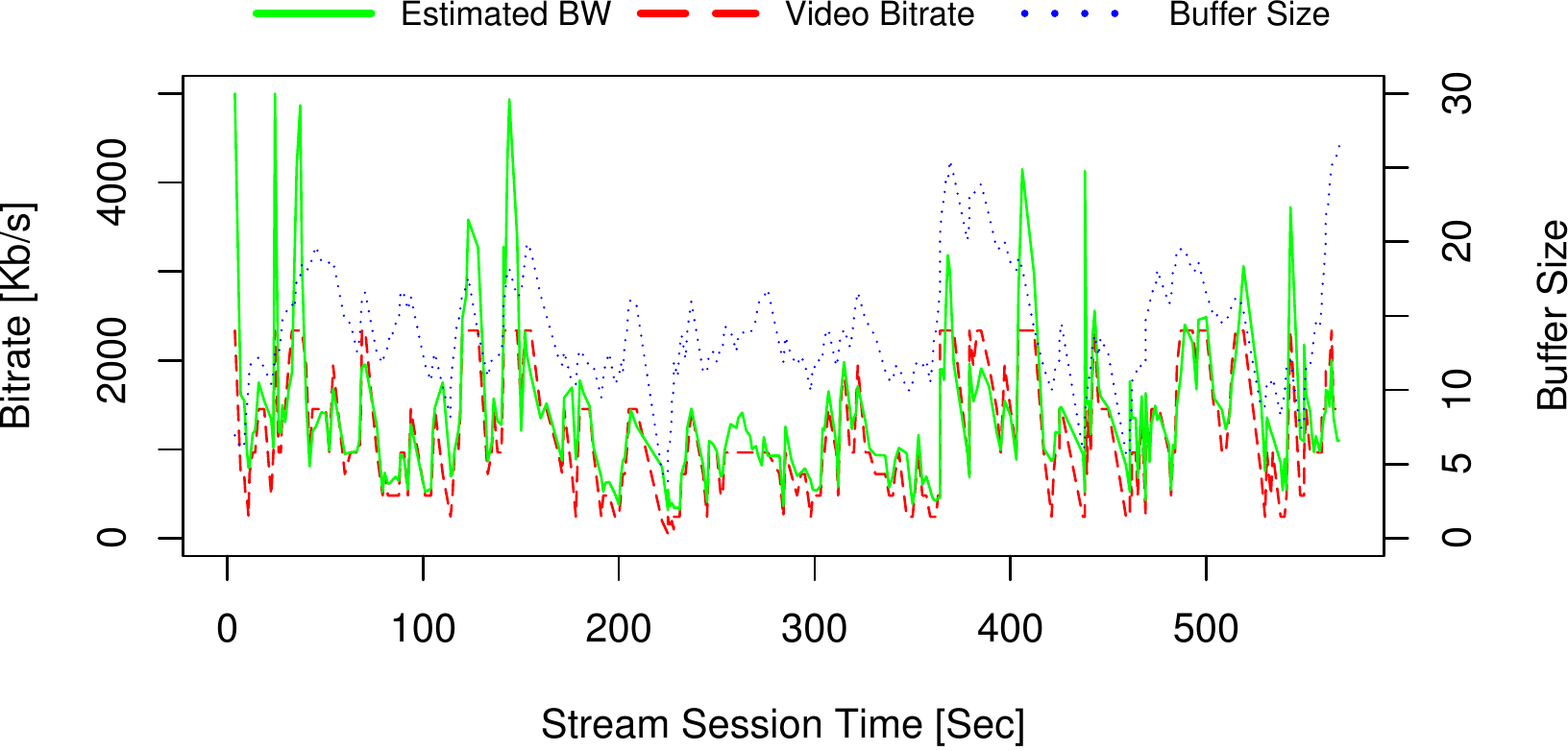}}
\subfigure[MAL]{\label{fig:i110HAMLbandwidthEtsimation}\includegraphics[width=0.3\textwidth]{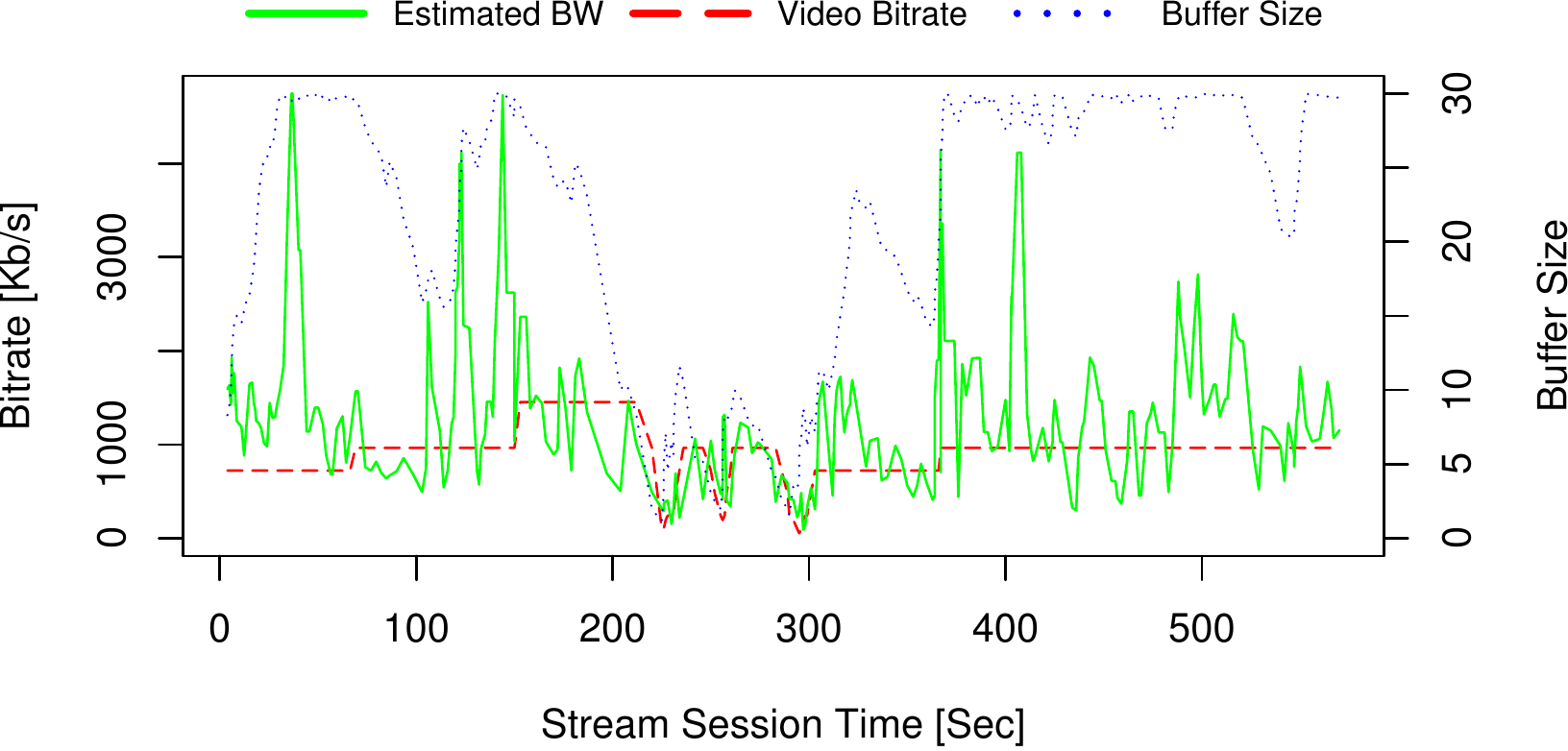}}
\subfigure[Geo-MAL]{\label{fig:i110GMALbandwidthEtsimation}\includegraphics[width=0.3\textwidth]{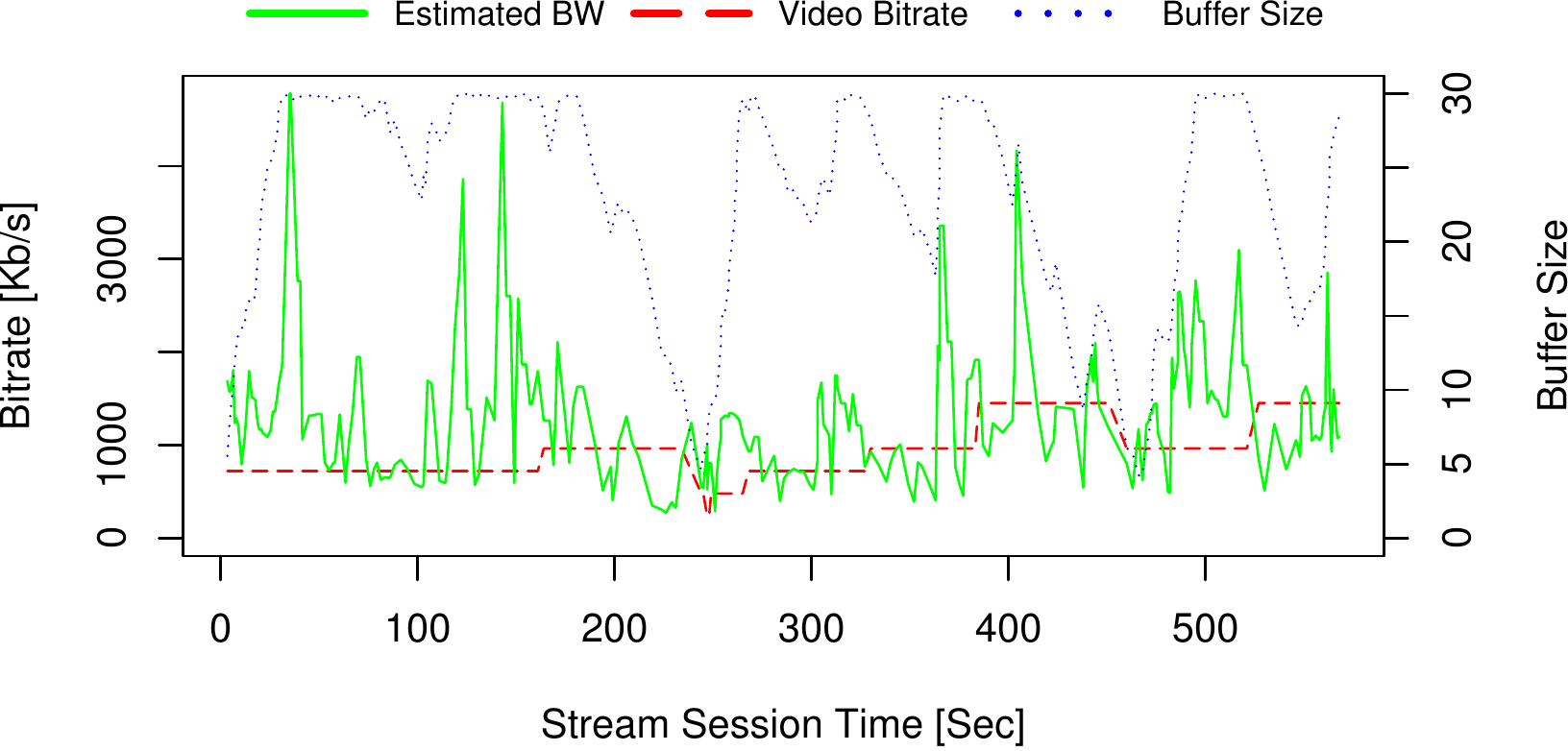}}
\subfigure[1-Predict]{\label{fig:i110PredictbandwidthEtsimation}\includegraphics[width=0.3\textwidth]{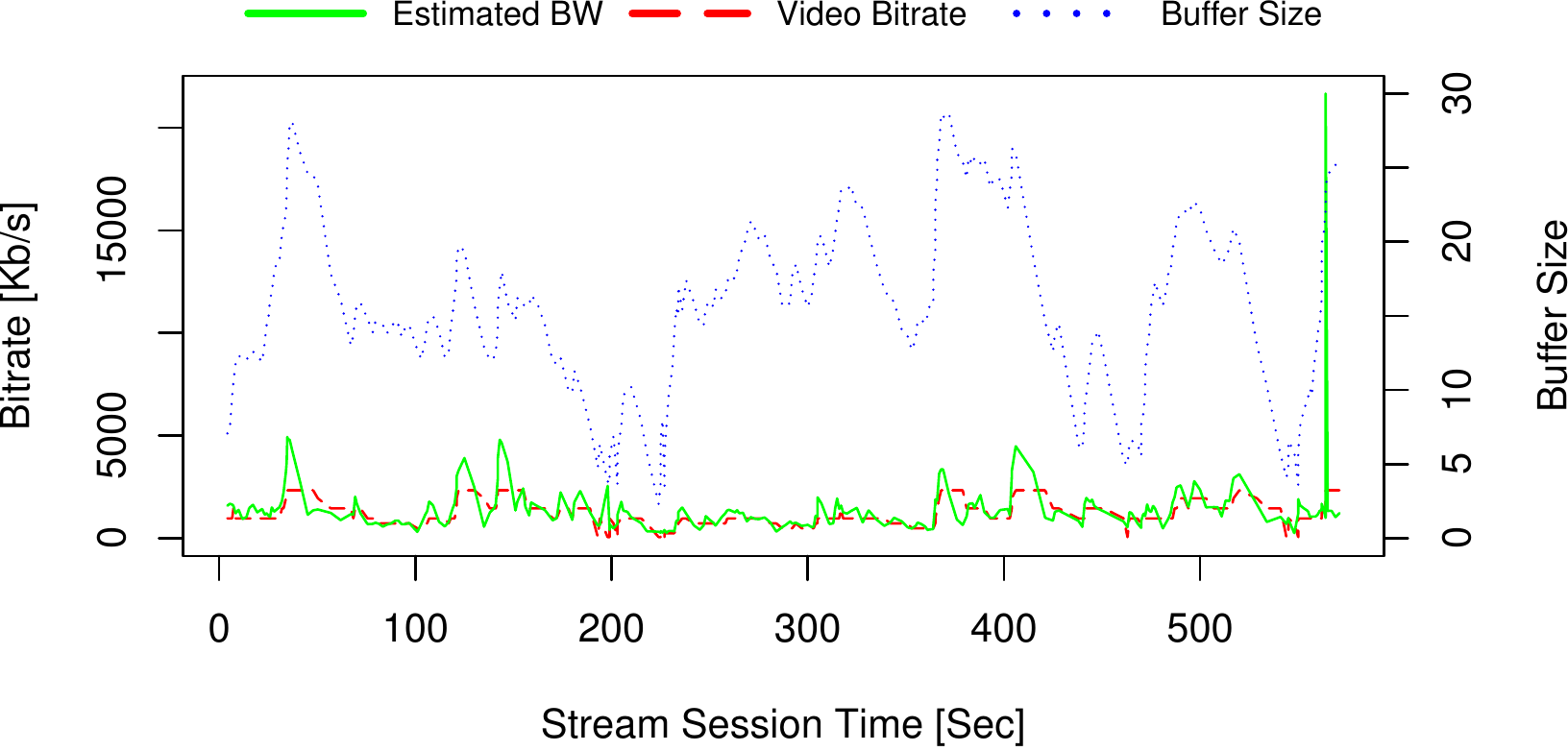}}
\subfigure[n-Predict]{\label{fig:i110NPredictbandwidthEtsimation}\includegraphics[width=0.3\textwidth]{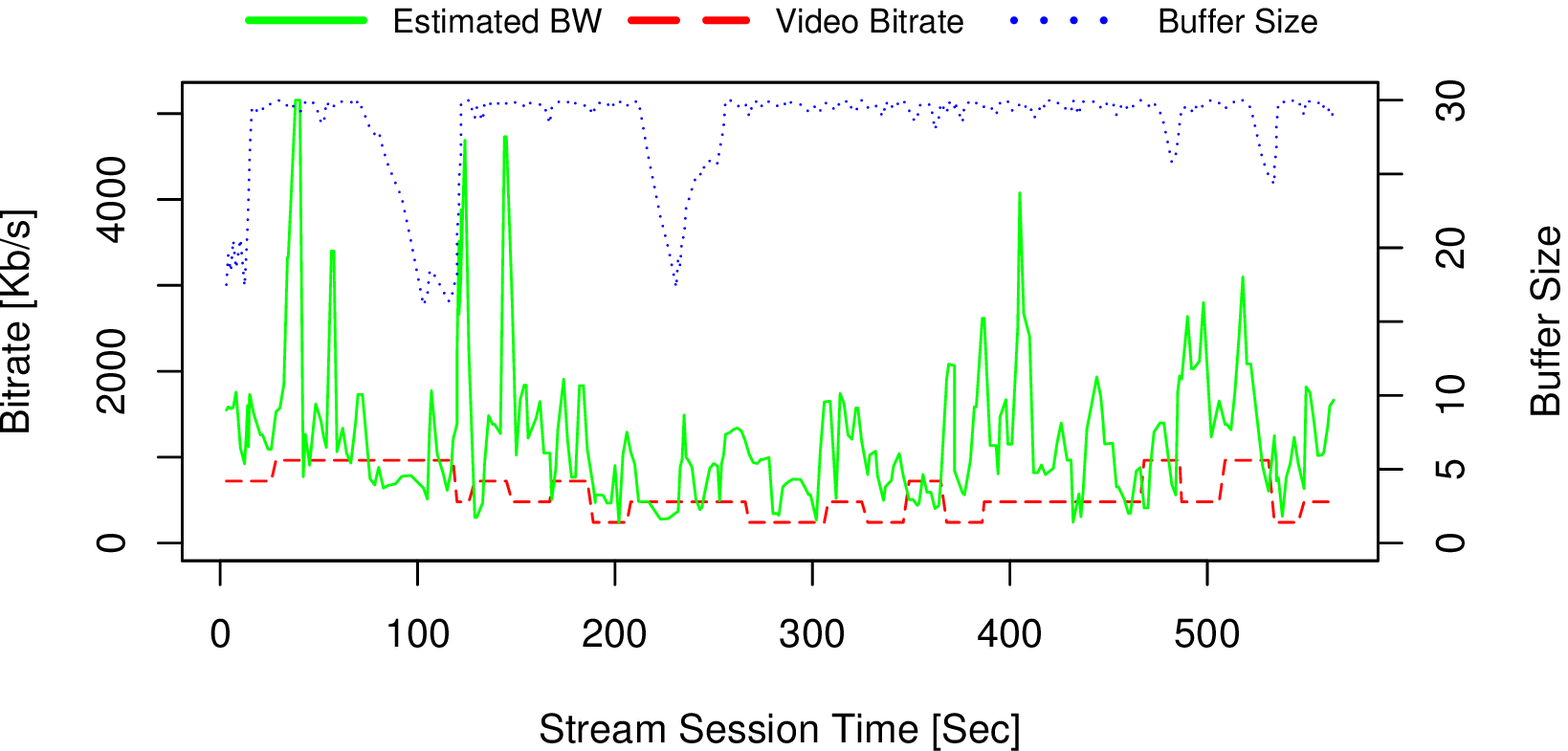}}
\subfigure[PBA]{\label{fig:i110HotmobilebandwidthEtsimation}\includegraphics[width=0.3\textwidth]{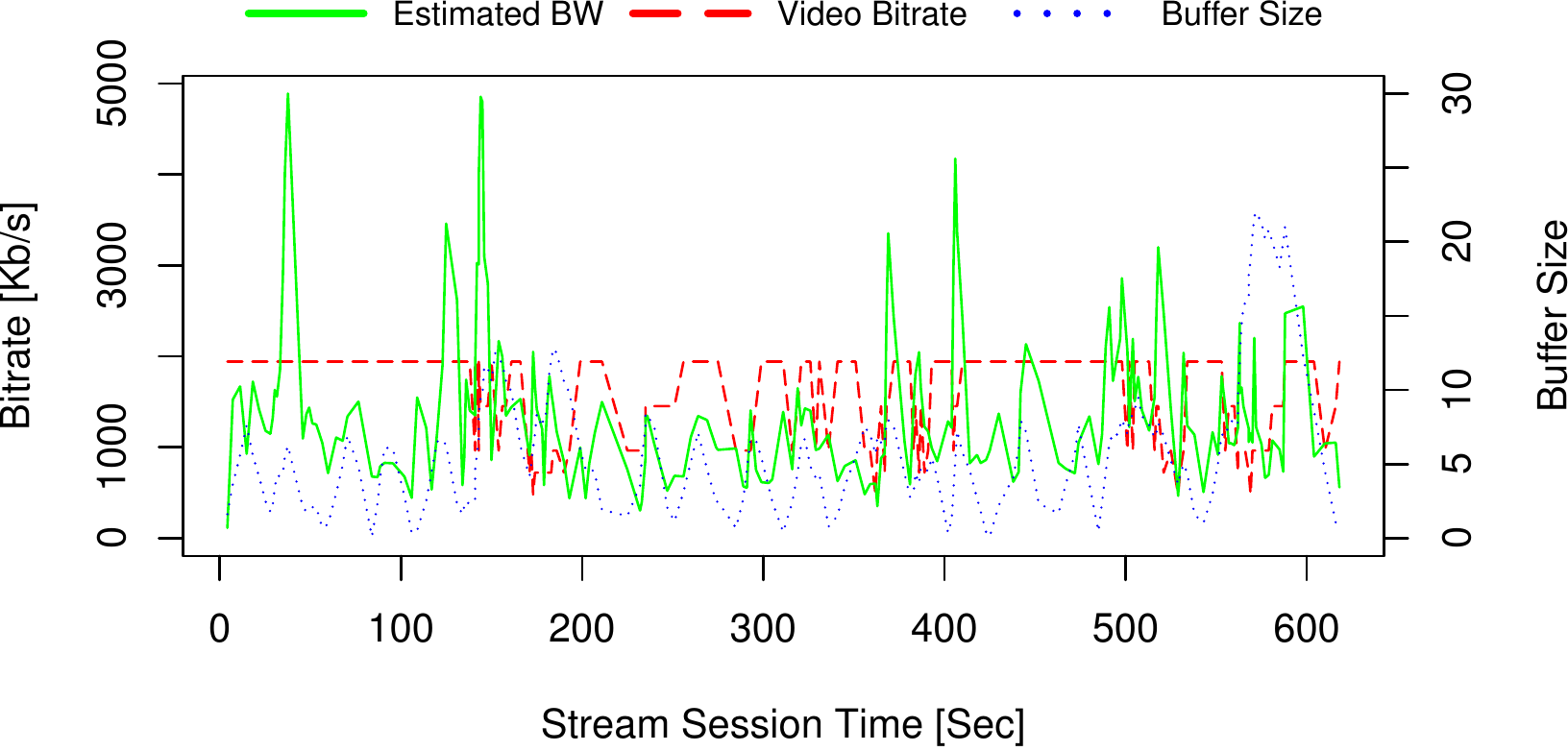}}
\subfigure[MASERATI]{\label{fig:i110MaseratibandwidthEtsimation}\includegraphics[width=0.3\textwidth]{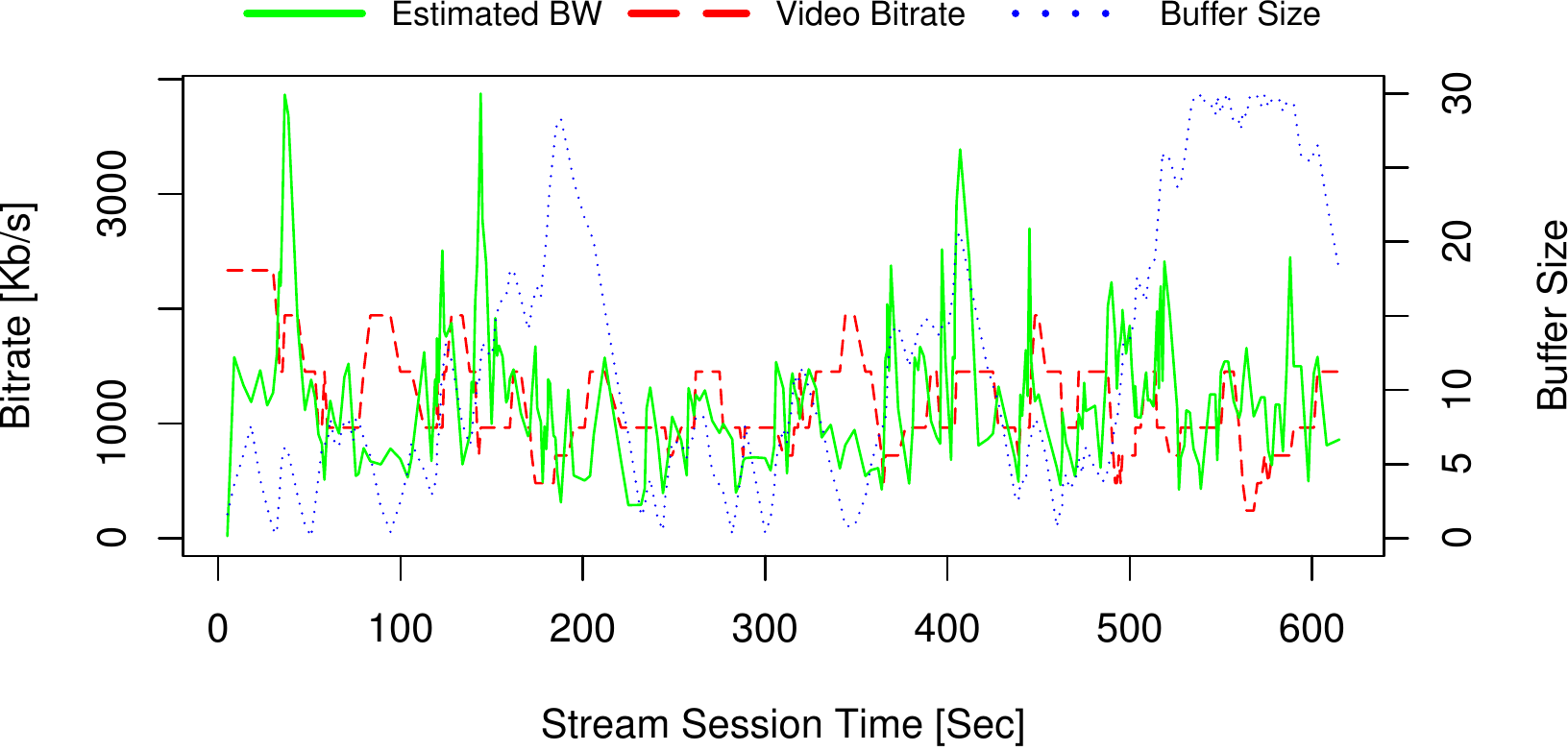}}
\caption{Algorithms' bandwidth and buffer estimates with the selected
  video bitrate for the Interstate I110 path.}
\label{fig:I110AtestResultComparison}
\end{figure*}

In this section we present the experimental results for Interstate
I110. Fig.\ref{fig:I110AtestResultComparison} presents the bandwidth
estimate, the downloaded bitrate and the buffer estimate.

Fig. \ref{fig:i110MaxBwbandwidthEtsimation} shows that MaxBW had a
relatively high bandwidth. Nevertheless, the algorithm selected the
suitable representation. MaxBW does not constrain the number of
representation switches. As a result, the algorithm had the highest
number of quality switches ($187$). MaxBW achieved the highest eMOS
score compared to all non-crowd algorithms without any re-buffering
events (see Table \ref{tab:Average_eMOS}). It is noteworthy that the
eMOS score includes factors such as the number of switch events and
re-buffering events.

Fig. \ref{fig:i110HAMLbandwidthEtsimation} shows that MAL smoothed
most of the small bit rate variations and gave restrained bandwidth
estimates that translated into a low number of quality switches
($33$).  It did so without any prior knowledge about the path network
conditions. Note that the algorithm adjusted too late to the
decreasing channel throughput ($150-200$ seconds). When the channel
bitrate decreased even further, the algorithm failed to recognize it
and encountered $4$ short re-buffering events.

Hao et al.\cite{hao2014gtube} suggested two algorithms: 1-predict and
n-predict. Their goal was to put forward a crowd-based algorithm with
fewer representation switches that could also minimize re-buffering
events. Fig. \ref{fig:i110PredictbandwidthEtsimation} shows that
1-predict's bit rate estimates was relatively high. Even though the
number of representation switches was high ($95$) the algorithm
maintained a balanced playout buffer and downloaded high quality
segments without re-buffering events. The n-predict algorithm achieved
very different results compared to the 1-predict
algorithm. Fig. \ref{fig:i110NPredictbandwidthEtsimation} shows that
the n-predict algorithm had the lowest number of representation
switches and that the playout buffers were extremely high. But the
n-predict algorithm tended to select lower representations; thus, its
eMOS score is the lowest (see Table \ref{tab:Average_eMOS}).

The Prediction Based Adaptation (PBA) \cite{Zou2015} algorithm
considers the buffer occupancy based on three buffer thresholds and
aggressively tries to stabilize rate selection.
Fig. \ref{fig:i110HotmobilebandwidthEtsimation} shows that the
algorithm tried to stay in the same representation but the buffer
occupancy tended to fluctuate which caused serious re-buffering
events.

The MASERATI algorithm uses a weighted average which considers the
past estimates and the adjusted bandwidth from the crowd
database. Fig. \ref{fig:i110MaseratibandwidthEtsimation} shows that
there was a relatively high number of representation switches ($82$)
whereas the total re-buffering duration was $39$ seconds. For a crowd
algorithm the average estimated throughput was low compared to other
crowd algorithms.

The Geo-MAL algorithm increased the number of representation switches
and minimized the number of re-buffering
events. Fig. \ref{fig:i110GMALbandwidthEtsimation} and
Table. \ref{tab:Average_eMOS} show that Geo-MAL improved the MAL eMOS
score by $9.6\%$ without any re-buffering events. The MAL algorithm
was designed for multicast networks.  Thus, its throughput estimates
are low and it tends to download lower representations. It tends to
smooth the network's fluctuations and thus its bandwidth estimates are
relatively low.

The Geo-MaxBW (Fig. \ref{fig:i110GeoMaxBWbandwidthEtsimation}) crowd
adaptation replaces the MaxBW's previous segment throughput estimate with
the crowd throughput estimate. The result was a slight reduction in
the number of representation switches and the average eMOS score
increased by $3.3 \%$.

Our GPAL algorithm is a buffer based approach inspired by the MaxBW
algorithm combined with crowd
knowledge. Fig. \ref{fig:i110GPALbandwidthEtsimation} shows that the
algorithm utilized the crowd in the most effective way and yielded an average
throughput estimate of $1.3 \lbrack Mb/s\rbrack$ compared to the other
crowd algorithms that generated lower utilization. GPAL had a relatively 
low number of representation switches ($48 \%$). Table
\ref{tab:Average_eMOS} shows that the algorithm had the best average
eMOS score.

\subsection{Interstate I405 Results}
\label{Interstate I405 Results}

\begin{figure*}[htbp]
\centering
\subfigure[GPAL]{\label{fig:i405GPALbandwidthEtsimation}\includegraphics[width=0.3\textwidth]{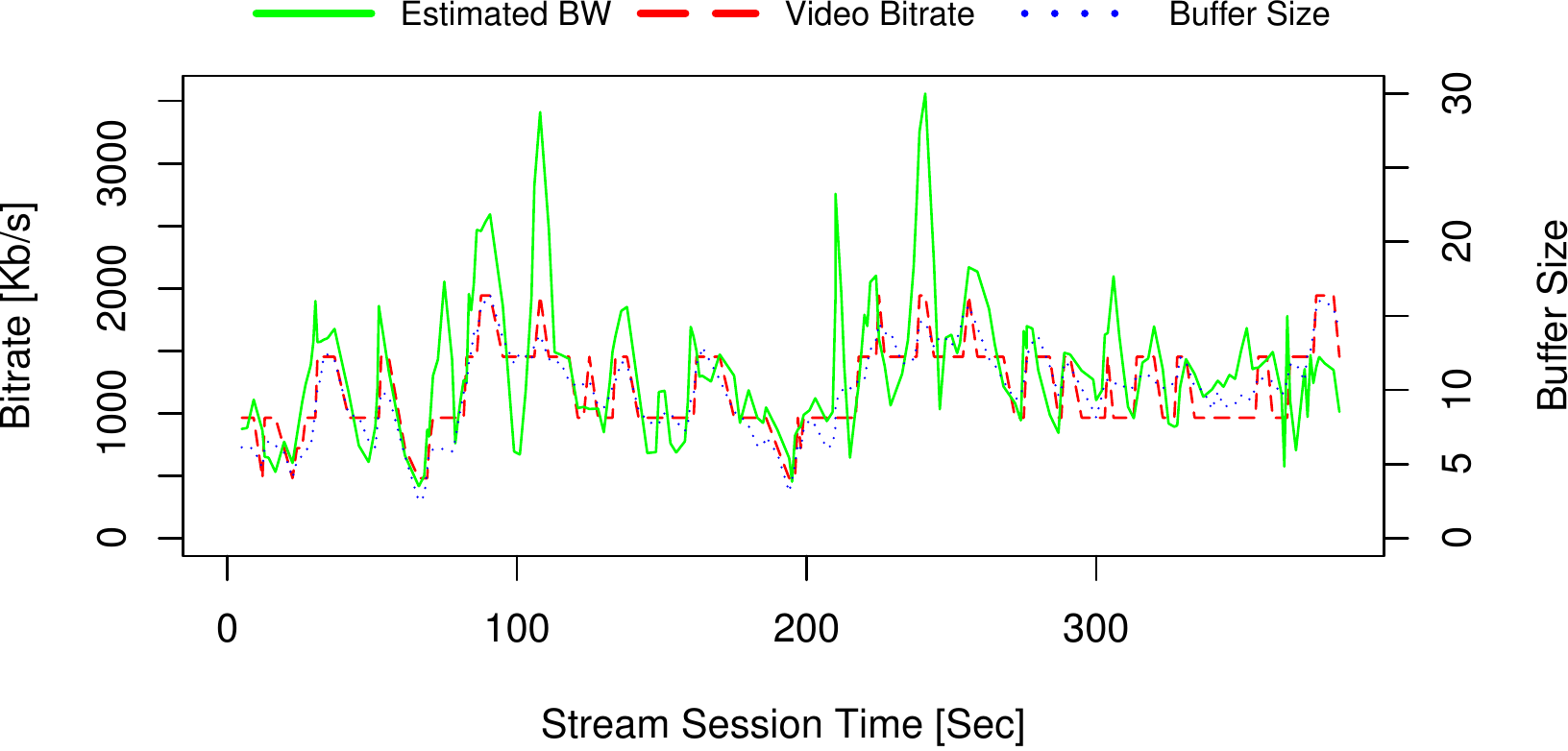}}
\subfigure[MaxBW]{\label{fig:i405MaxBwbandwidthEtsimation}\includegraphics[width=0.3\textwidth]{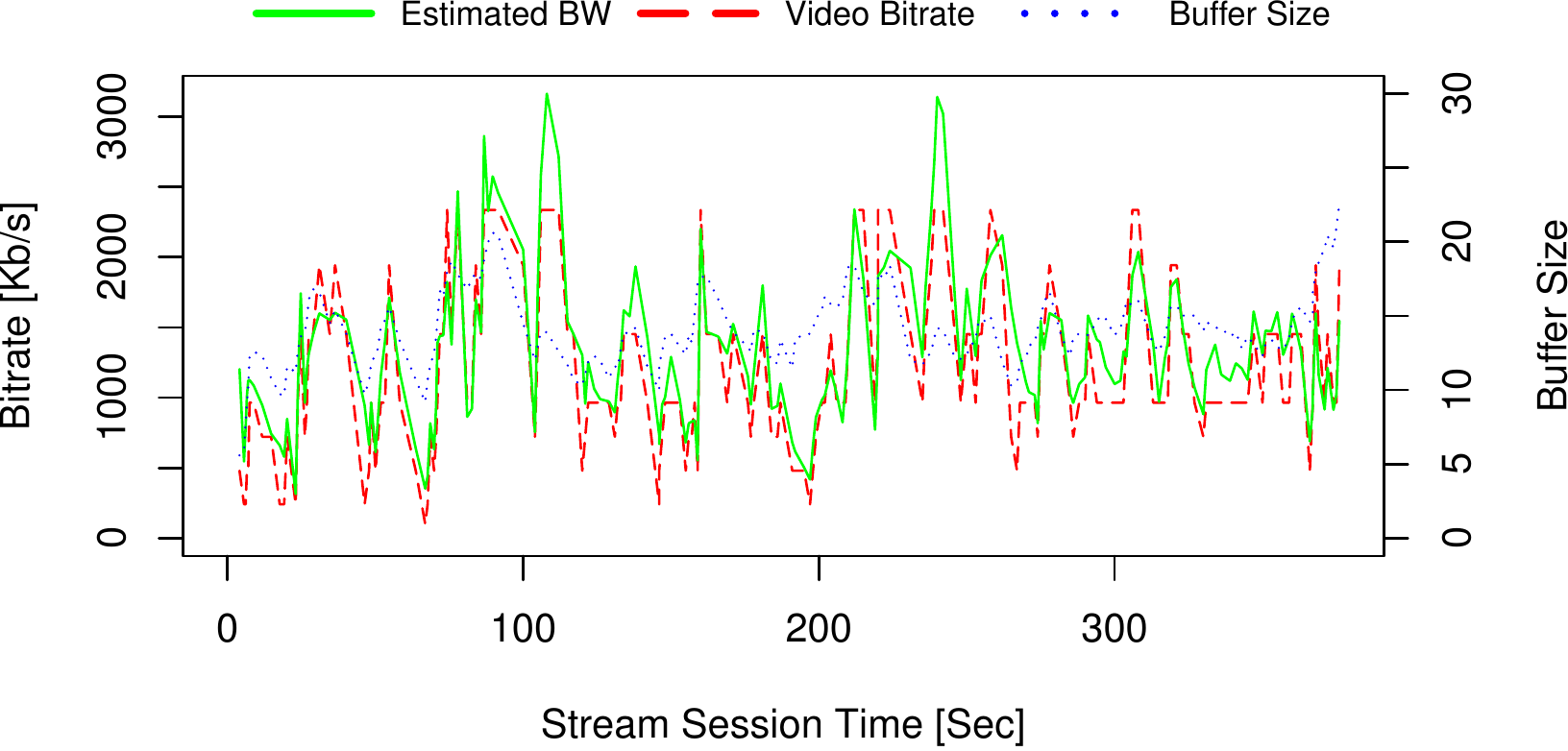}}
\subfigure[Geo-MaxBW]{\label{fig:i405GPALbufferEtsimation}\includegraphics[width=0.3\textwidth]{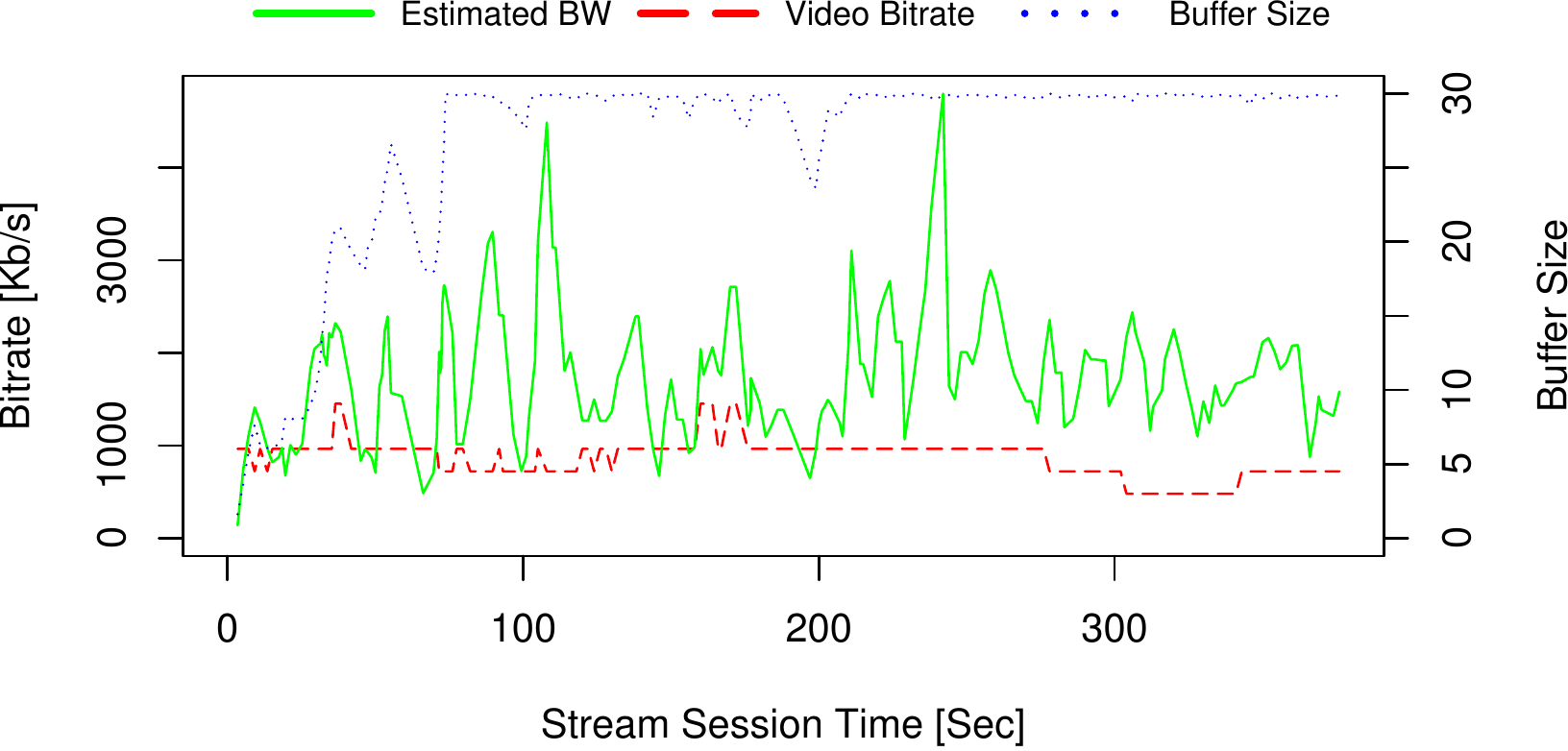}}
\subfigure[MAL]{\label{fig:i405HAMLbandwidthEtsimation}\includegraphics[width=0.3\textwidth]{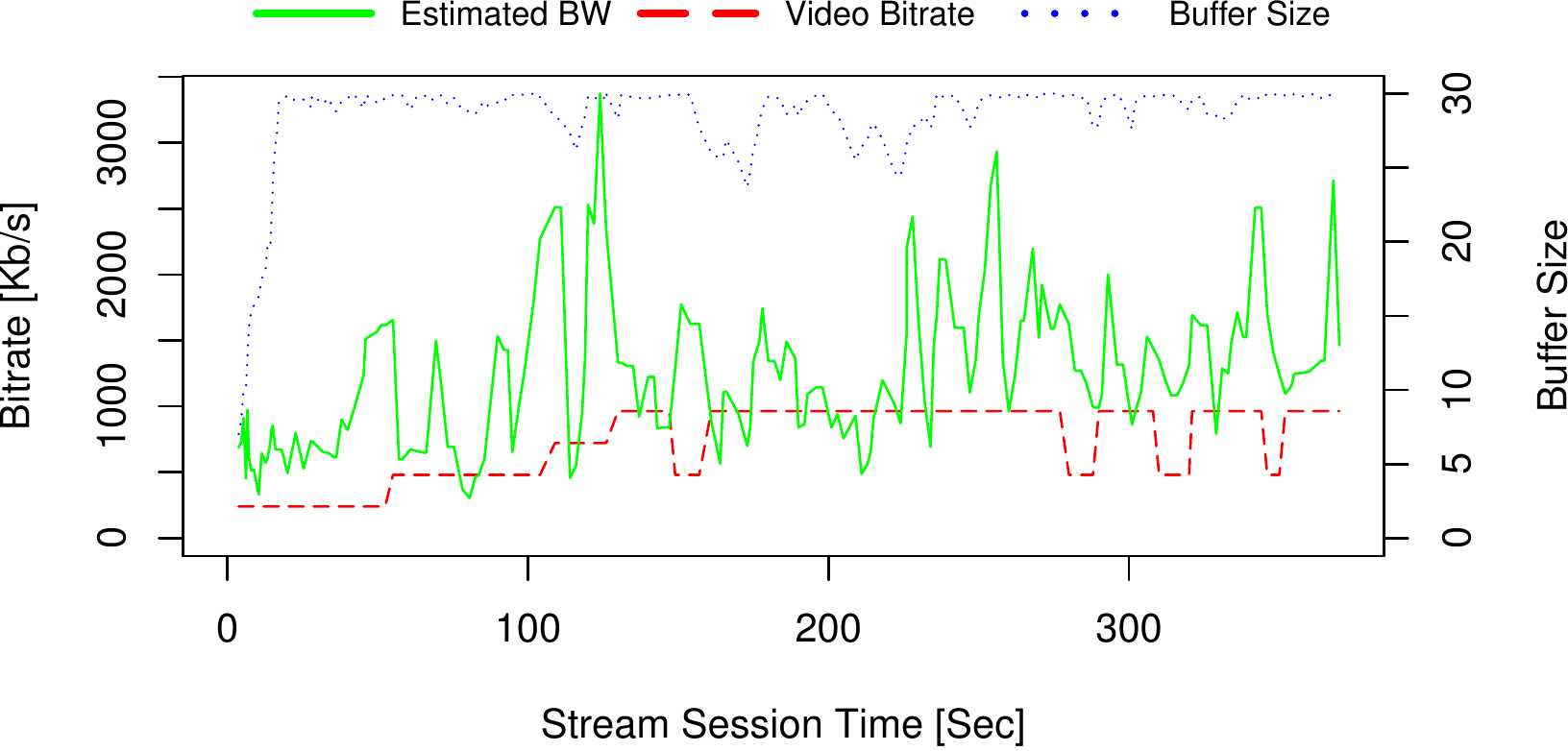}}
\subfigure[Geo-MAL]{\label{fig:i405GMALbandwidthEtsimation}\includegraphics[width=0.3\textwidth]{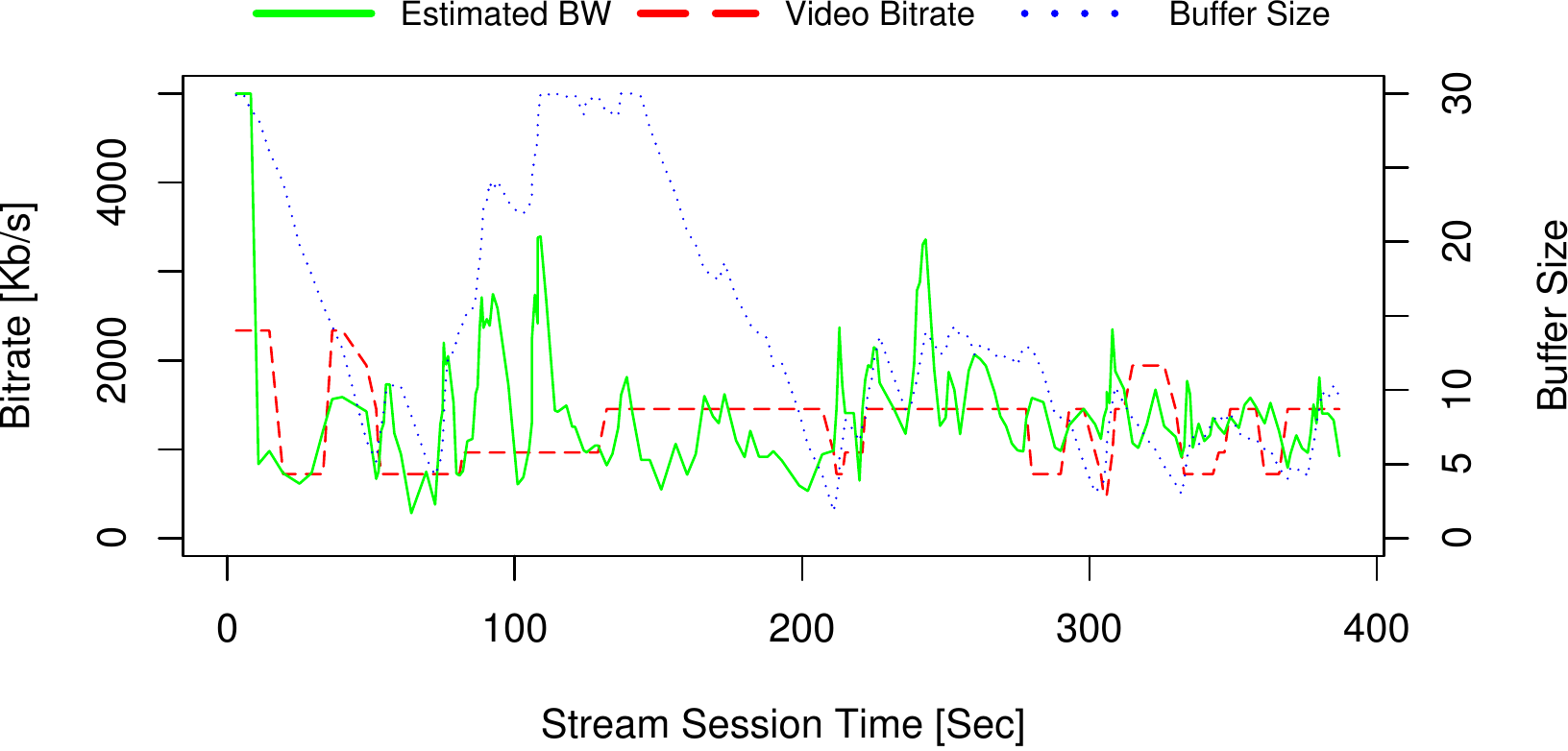}}
\subfigure[1-Predict]{\label{fig:i405PredictbandwidthEtsimation}\includegraphics[width=0.3\textwidth]{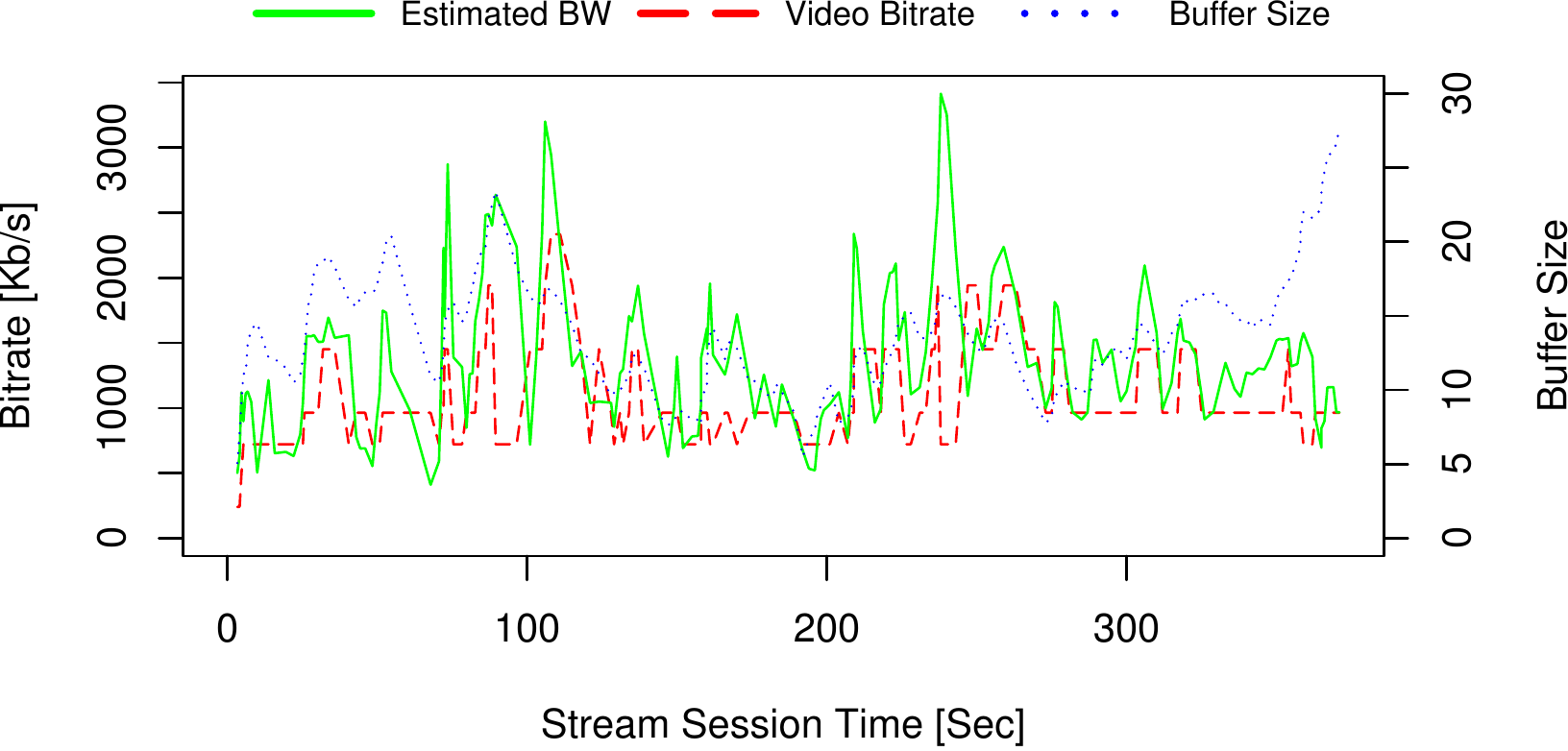}}
\subfigure[n-Predict]{\label{fig:i405NPredictbandwidthEtsimation}\includegraphics[width=0.3\textwidth]{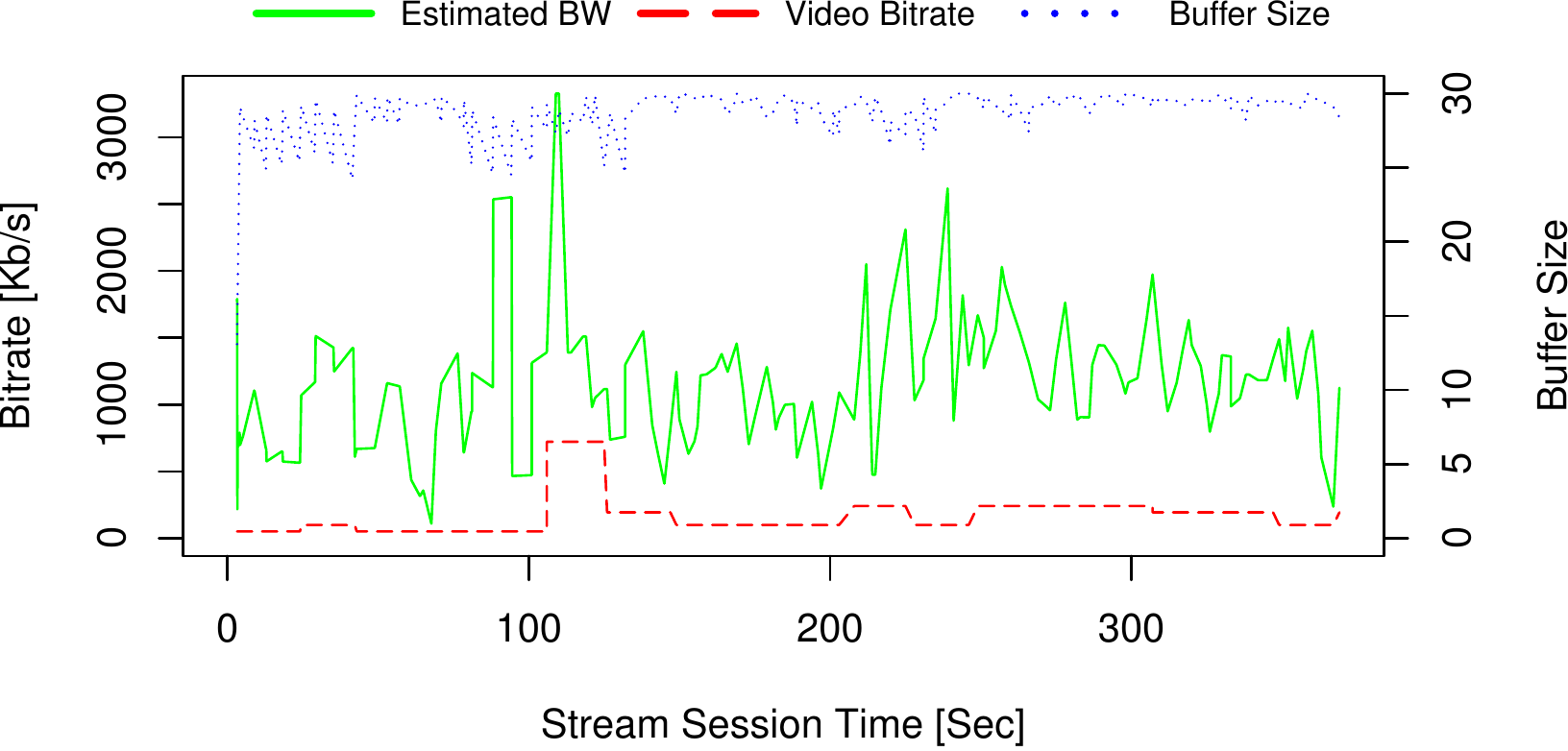}}
\subfigure[PBA]{\label{fig:i405HotmobilebandwidthEtsimation}\includegraphics[width=0.3\textwidth]{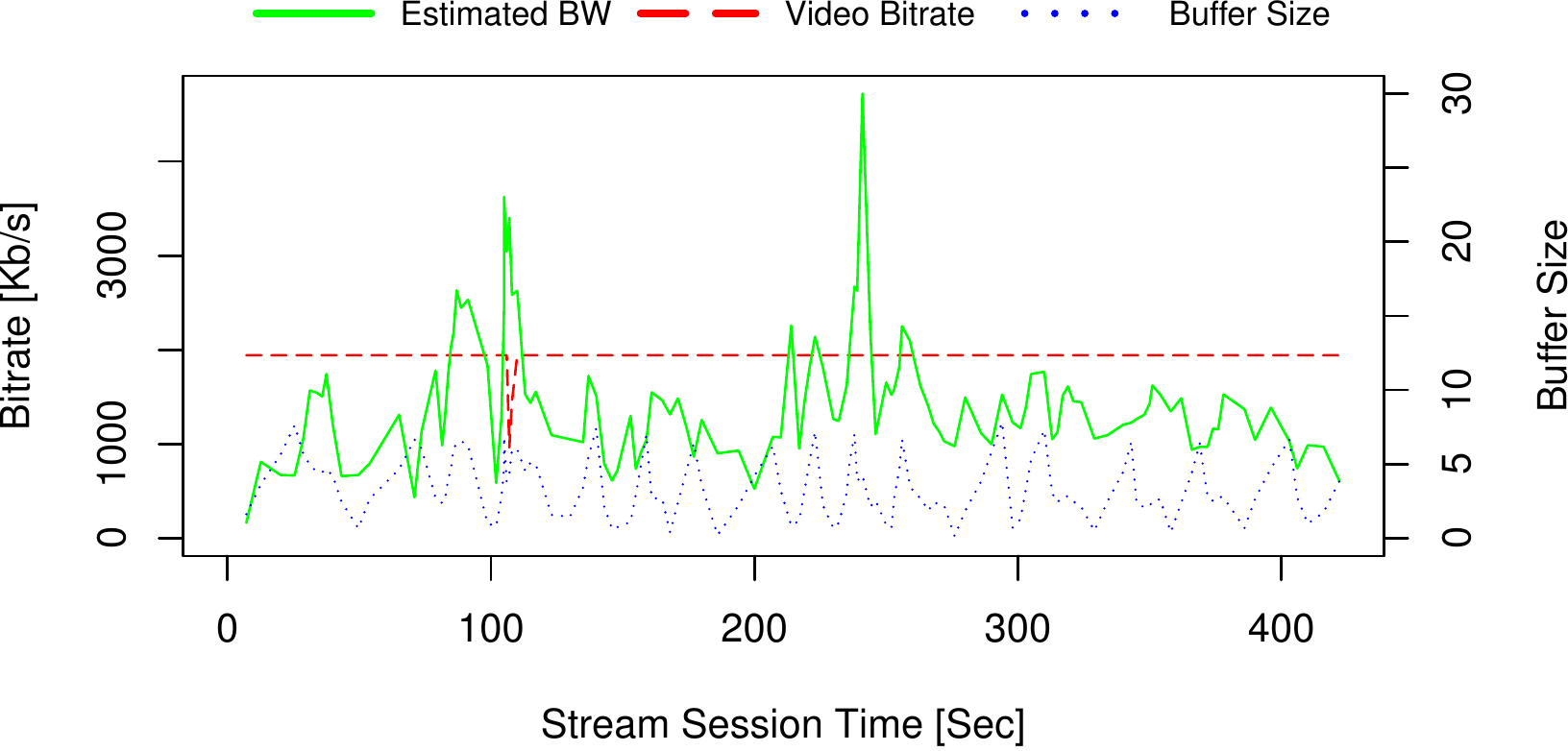}} 
\subfigure[MASERATI]{\label{fig:i405MaseratibandwidthEtsimation}\includegraphics[width=0.3\textwidth]{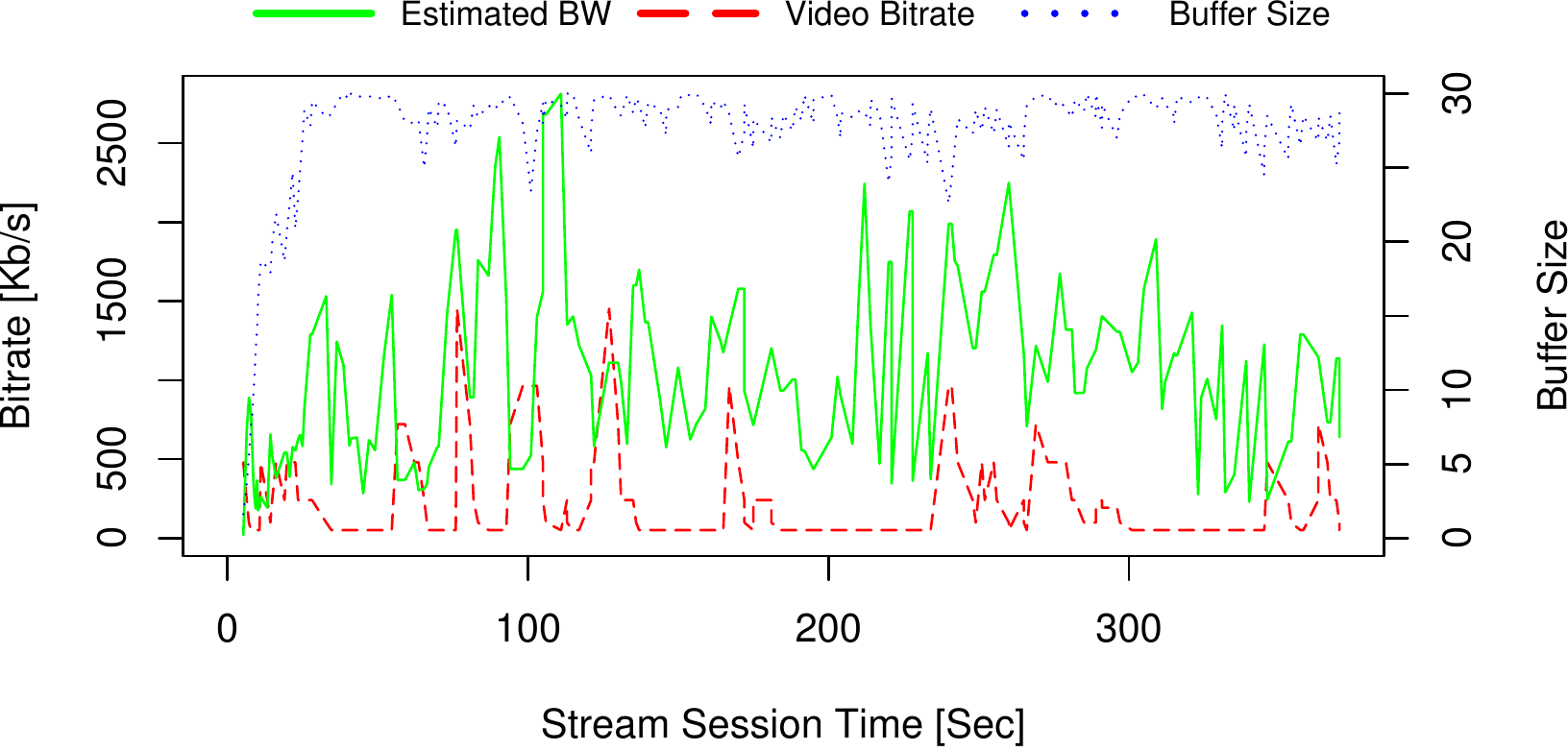}}
\caption{Algorithms' bandwidth and buffer estimates with the selected
  video bitrate for the Interstate I405 path.}
\label{fig:I405AtestResultComparison}
\end{figure*}

Interstate I405 (see Fig. \ref{fig:i405Info}) differs from
Interstate I110 and its bandwidth is spread better in the path. This
caused the algorithms to select different representations than for
I110. Fig. \ref{fig:I405AtestResultComparison} shows that the MAL
algorithm, which is an exponential moving average based algorithm,
smooths the bandwidth estimate. Surprisingly, compared to I110 the
algorithm did not have re-buffering events, but this time Geo-MAL
did. MAL based algorithms do not check whether the estimated throughput or
crowd throughput are lower than the selected representation. This
behavior led to re-buffering. The MaxBW algorithm exhibited good
performance, similar to I110. However the number of representation
switches increased to $130$. The Geo-MaxBW reduced the number of
representation switches to $28$. The other algorithms' behavior was
similar to I110.  Table \ref{tab:Average_eMOS} shows that the average
eMOS score for MaxBW gave the best result for non-crowd algorithms
whereas GPAL outperformed all the other algorithms.

\section{Conclusion}
\label{Conclusions}

We showed that the use of real-world crowd data can improve existing
algorithms and demonstrated it on two different algorithms: Geo-MAL
and Geo-MaxBW. Geo-MAL presented a $9.6 \%$ average eMOS score
improvement over MAL whereas Geo-MaxBW had a $3.3 \%$ improvement over
MaxBW. We proposed a new crowd-based algorithm called GPAL that
outperformed all other state-of-the-art algorithms.  We conclude that
an optimal adaptation logic should estimate the distance between the
current conditions and the cloud conditions. Our future work will aim
to design an adaptation algorithm that can leverage the advantages of
past download algorithms with crowd knowledge based on the conclusions
drawn from this work. An interesting approach would be to implement
machine learning algorithms (similar to Claeys et
al. \cite{claeys2014design}) combined with crowd data. An additional
interesting research direction would be to harness a client-side
pre-fetch and $HTTP2$ server-side push mechanism based on crowd
knowledge.

\bibliographystyle{unsrt}
\bibliography{references}

\end{document}